\documentclass[preprint,12pt]{elsarticle}

\usepackage{graphicx}
\usepackage{amssymb,mathtools}
\usepackage{epstopdf}
\usepackage{bm}
\usepackage{color}

\usepackage{bm,amsmath,amssymb}

\long\def\comment#1{ }

\newcommand{\beq}{\begin{eqnarray}}
\newcommand{\eeq}{\end{eqnarray}}
\newcommand{\nn}{\nonumber\\}

\newcommand{\rmd}{{\rm d}}
\newcommand{\rme}{{\rm e}}

\newcommand{\del}{\partial}

\newcommand{\mcal}{\mathcal}

\newcommand{\F}{\mcal{F}}

\begin{document}

%%%%%%%%%%%%%%%%%%%%%%%%%%%%%%%%%%%%%%%%%%

\begin{frontmatter}

\title{\Large Energy flow along the medium-induced parton cascade}

\author{J.-P. Blaizot }

\address{Institut de Physique Th\'{e}orique de Saclay, CNRS/URA2306, CEA Saclay,
F-91191 Gif-sur-Yvette, France}
\author{Y. Mehtar-Tani}
\address{Institute for Nuclear Theory, University of Washington, Seattle, WA 98195-1550, USA}

\begin{abstract}

We discuss the dynamics of parton cascades that develop in dense QCD matter, and contrast their properties with those of similar cascades of gluon radiation in vacuum. We argue that such cascades belong to two distinct classes that are characterized respectively by an increasing or a constant (or decreasing) branching rate along the cascade. In the former class, of which the BDMPS, medium-induced, cascade constitutes a typical example, it takes a finite time to transport a finite amount of energy to very soft quanta, while this time is essentially infinite in the latter case, to which the DGLAP cascade belongs. The medium induced cascade  is accompanied  by a constant flow of energy towards arbitrary soft modes, leading eventually to the accumulation of the initial energy of the leading particle at zero energy. It also exhibits scaling properties akin to wave turbulence. These properties do not show up  in the cascade that develops in vacuum. There,  the energy accumulates in the spectrum at smaller and smaller energy as the cascade develops, but the energy never flows all the way down to zero energy. Our analysis suggests that the way the energy is shared among the offsprings of a splitting gluon has little impact on the qualitative properties of the cascades, provided the kernel that governs the splittings is not too singular.  

\end{abstract}
\begin{keyword}
Perturbative QCD, Heavy Ion Collisions, Jets, QCD cascades\\
\end{keyword}

\end{frontmatter}
\begin{flushleft}
\footnotesize{PACS numbers: 12.38.-t,24.85.+p,25.75.-q}
\end{flushleft}
\begin{flushright}
INT-PUB-15-002
\end{flushright}

\section{Introduction}
\label{sec:intro}
The strongly collimated cascades of gluons that are radiated by an energetic parton is an essential ingredient of the physics of jets that are produced in various high energy collisions, including heavy ion collisions (for a recent review see \cite{Mehtar-Tani:2013pia} and references therein). This is because the inclusive distribution of the final particles is to a large extent determined by that of the radiated gluons at the time of hadronization, a feature known as the local parton-hadron duality \cite{Dokshitzer:1991wu}. In the case of heavy ion collisions, it is expected that multiple interactions of the hard partons with the deconfined matter present in the final state,   induce another type of cascade \cite{Baier:2000sb,Jeon:2003gi,Blaizot:2013hx}. Its elementary radiation process is described by the BDMPS-Z theory \cite{Baier:1996kr,Baier:1996sk,Baier:1998kq,Zakharov:1996fv,Wiedemann:2000za,Arnold:2002ja}. In the ideal case where only medium induced processes are retained, the resulting cascade exhibits particular features that make it very different from the cascade of gluons generated by radiation in vacuum,  commonly described by the DGLAP evolution equation \cite{Gribov:1972ri,Altarelli:1977zs,Dokshitzer:1977sg}. In particular, the BDMPS cascade exhibits scaling properties, akin to wave turbulence \cite{Nazarenko}, that are apparently absent in the DGLAP cascade. Also, as was recently emphasized, the BDMPS cascade provides a natural mechanism for the transport of energy towards large angles, which contrasts with the strong angular ordering of QCD cascades in vacuum \cite{Mehtar-Tani:2014yea,Blaizot:2014ula,Kurkela:2014tla,Blaizot:2014rla}.

The main motivation of the present paper is to analyze the origin of the qualitative differences between the two types of cascades, with the goal of fostering our understanding of the BDMPS cascade. This cascade, as we have just recalled, may play an important role in determining the properties of jets produced in ultra relativistic heavy ion collisions. Insight will be obtained by comparing its properties with those of the more familiar DGLAP cascade.  As we shall see, a quantity that plays a crucial role in this comparison is the characteristic  time,  $t_\ast(\omega)$, that  it takes a gluon of energy $\omega$ to split into two other gluons. We refer to this time as the branching time.   In the BDMPS cascade the branching time is an increasing function of  $\omega$, $t_\ast(\omega)\sim \sqrt{\omega}$, so that it decreases along the cascade as gluons of smaller and smaller energies are emitted. In other words, the branchings accelerate, and as a result it takes a finite time to transport a finite amount of energy from the leading particle to gluons carrying vanishingly small amounts of energy. This time, which we shall refer to as the stopping time \cite{Arnold:2009ik}, is infinite for the DGLAP cascade in which $t_\ast(\omega)$ is   independent of $\omega$.  

Further features of the cascades follow from this crucial difference in the branching times. Consider for instance the energy distribution $D(\omega)$. In the  BDMPS cascade  this exhibits a persistent scaling behavior all the way down to $\omega=0$, of the form $D(\omega)\sim 1/\sqrt{\omega}$, while in the DGLAP cascade  scaling cannot be achieved in a finite time. The emergence of such a scaling relates to the existence of a stationary solution to the equation that governs the evolution of the energy distribution as a function of time. This stationary solution is of the form
\beq
D_\text{st}(\omega)=\frac{t_\ast(\omega)}{\omega},
\eeq 
and is associated to a constant (independent of $\omega$) flow of energy,  
\beq
{\cal F}(\omega)\equiv \frac{\del {\cal E}(\omega)}{\del t}\sim \frac{\omega D_\text{st}(\omega )}{t_\ast(\omega)}=\text{const.}
\eeq
where ${\cal E}(\omega)$ stands for the total energy carried by gluons with energies larger than $\omega$, i.e., ${\cal E}(\omega)=\int_\omega\rmd \omega' D(\omega').$ 
The existence of  a constant flow of energy at arbitrarily small $\omega$ implies that  
\beq
{\cal E}(\omega_0)=\int_{\omega_0} \rmd\omega D_\text{st}(\omega)=\int_{\omega_0} \frac{\rmd\omega}{\omega} t_\ast(\omega)
\eeq 
remains finite when $\omega_0$ tends to zero. This forces  $t_\ast(\omega)$ to be an increasing function of $\omega$, which is indeed the case for the BDMPS cascade, for which $t_\ast(\omega)\sim \sqrt{\omega}$.  In the DGLAP cascade $t_\ast(\omega)$ is constant, i.e. independent of $\omega$, and the integral of the stationary scaling spectrum diverges. As we shall see, a scaling spectrum $D(\omega)\sim 1/\omega $ would also be expected in this case, but it takes an infinite time to develop.  

The properties of the cascades depend also on the splitting kernel, that is, on the way the energy is distributed between the offsprings during a splitting. However, this turns out to have a minor effect on the main characteristics of the cascade, as compared to that of the transport time scale just mentioned, at least as long as the splitting kernel is not too singular. In that case the cascades develop as if the branching were completely democratic, with the two offsprings taking each half the energy of the parent gluon. The interactions responsible for the splittings can then be considered as local (in energy space). As already mentioned, these properties of the BDMPS cascade that we have briefly listed, are typical of wave turbulence \cite{Nazarenko}.

This paper is organized as follows. In the next section, we briefly recall the main features of the BDMPS-Z theory, and the properties of the ideal medium induced QCD cascade. By ideal we mean that we ignore the fact that the BDMPS mechanism is valid only within a range of energies of the emitting gluons, while we shall allow gluons to split all the way down to zero energy according to the BDMPS mechanism. This introduces small, unphysical, but well identified effects, that play however no role in the present discussion. We contrast the resulting properties of this ideal cascade with that obtained by solving (a simplified version of) the DGLAP evolution equations. In Section 3, we analyze the two types of cascades from a more general point of view. We isolate in the splitting kernel a contribution that we associate with a characteristic branching rate, and a function that determines the probability that one of the offspring carries a given fraction $z$ of the parent gluon. We show that the main distinction between the two cascades comes from the branching rate, and in particular whether it increases with decreasing energy or not. 

This general analysis is completed in Section 4 by a study of explicit models, that can be solved analytically, and that provide insight into the emergence of the scaling solutions. In these models we examine the two situations of constant or accelerating branchings, first in the case of complete democratic branching where the offsprings carry a fraction $z=1/2$ of the parent energy at each splitting, and then in the case where the distribution of energy fractions $z$ is independent of $z$. This analysis confirms the general discussion of Section 4. The main conclusions of the paper are summarized in Section 5. The paper contains also four technical appendices. Appendix A presents the construction of the analytic solution of the BDMPS cascade both in the case of an infinite medium and that of finite medium, and  Appendix B presents the solution of the simplified DGLAP evolution equation that we use in the main text. Appendix C and D provide more details on the analytic solutions of the models that are discussed in Section 4. 

%%%%%%%%%%%%%%%%%%%%%%%%
\section{The ideal in-medium QCD cascade}
%%%%%%%%%%%%%%%%%%%%%
\begin{figure}[ht]
   \center \includegraphics[width=11cm]{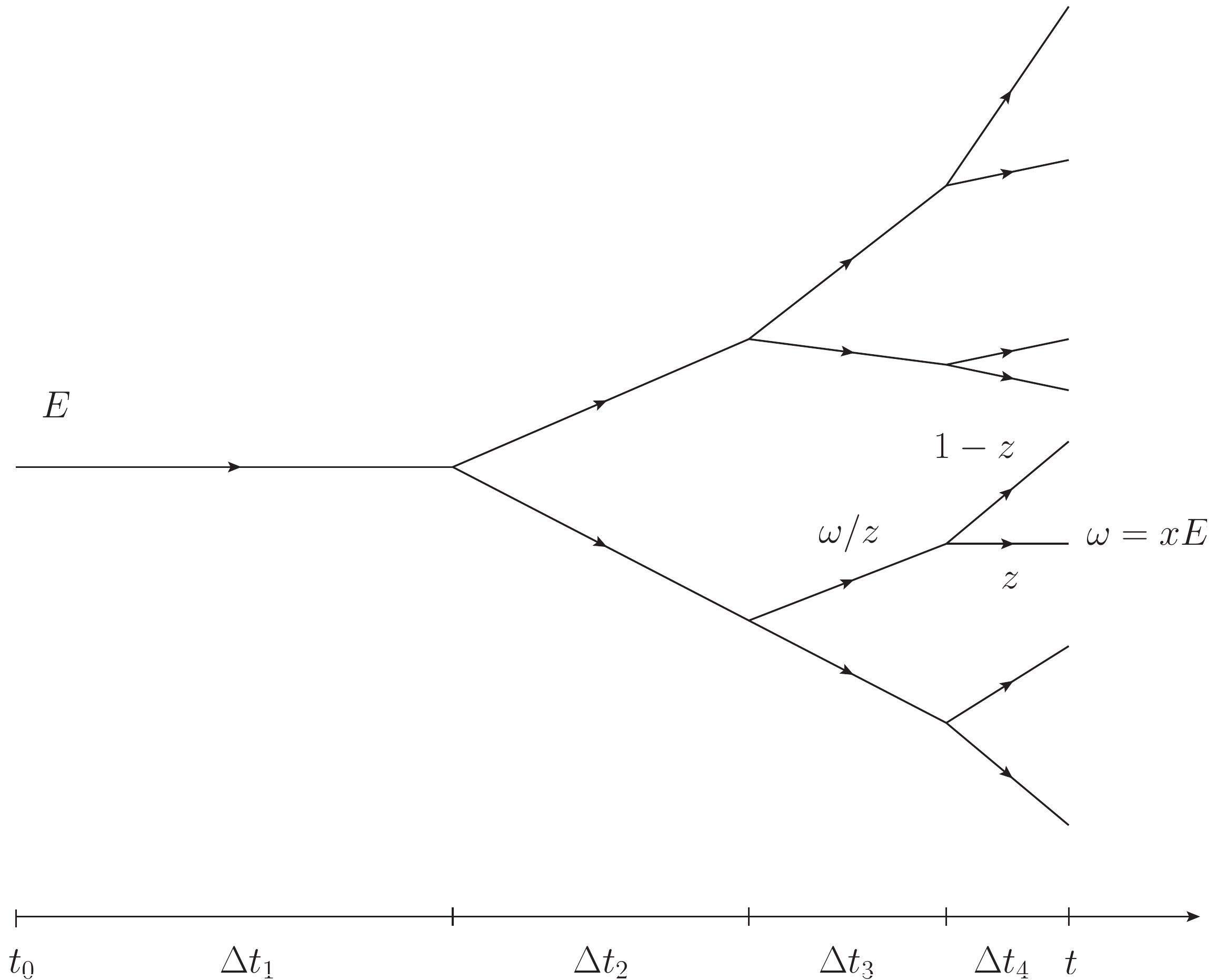} 
		\caption{Illustration of a gluon cascade that is initiated by a gluon with energy $E$. Four generations are displayed. The branching time $\Delta t_i$, that corresponds to the lifetime of generation $i$, decreases after each branching as in a BDMPS cascade. The inclusive distribution $D(x,t)$ measures the probability to find in the cascade, at time $t$,  a gluon with energy $xE$.  The rate equation (\ref{Dfin2}) describes how this distribution evolves with time $t$.}
		\label{fig0}
\end{figure}
We consider an energetic (on-shell) gluon, with initial energy $E$,  traveling through a medium, for instance a quark-gluon plasma\footnote{Our main results do not depend crucially on the detailed properties of the medium}. This gluon, referred to in the following as the {\em leading particle},  interacts with the plasma constituents, and these interactions induce radiation. The radiated gluons themselves radiate, eventually generating a cascade. Our goal is to study the average properties of such a cascade, focusing in this paper on the inclusive  gluon distribution, integrated over transverse momentum, 
\beq
D(x,t)\equiv  x \frac{\rmd N}{\rmd x},
\eeq
where  $x=\omega/E$ is the energy fraction of the gluon observed at some time $t$ along the cascade, cf. Fig~\ref{fig0} for an illustration. ($D(x,t)$ can be viewed as an energy density, with $D(x,t) \,\rmd x$ being the energy contained in modes\footnote{We use the term ``mode'' as a synonymous for radiated gluon throughout this paper} with energy fraction between $x$ and $x+\rmd x$.)  As was shown in \cite{Blaizot:2013vha},  $D(x,t)$ obeys the following equation\footnote{As originally derived, this  equation  is  written in terms of the light-cone time  $x^+=(t+z)/\sqrt{2}\approx t\sqrt{2}$, with $t$ the physical time. Similarly, its original kernel involves the component  $p_0^+\approx  p_0\sqrt{2}$ of the four momentum. Then $t_\ast$ is given by
$$
\frac{1}{t_\ast}\bar\alpha \sqrt{\frac{\hat q}{p^+_0}}\, x^+=\bar\alpha \sqrt{\frac{\hat q \sqrt{2}}{E}}t.
$$
Thus, modulo a redefinition of $\hat q$ ($\hat q\to \hat q\sqrt{2}$), one can reads Eq.~(\ref{Dfin2}) in terms of the physical  time $t$  and  energy $E$. This is the convention that we adopt throughout this paper.} 
 \beq\label{Dfin2}
\frac{\partial}{\partial t}D(x,t)&=&\frac{1}{t_\ast}\int_x^1 \rmd z\,  {\cal K}(z)\sqrt{\frac{z}{x}}
D\left(\frac{x}{z},t\right)
-\frac{1}{t_\ast}\int_0^1 \rmd z\,  {\cal K}(z)\frac{z}{\sqrt{x}}D\left(x,t\right),\nn
&\equiv&{\cal I}[D],
\eeq
where $t_\ast$ is a characteristic time scale (to be discussed shortly). The right hand side of Eq.~(\ref{Dfin2}), ${\cal I}[D]$, has the form of a `gain term' minus a `loss term'.  The gain term  describes the production of a new gluon with
energy fraction $x$  via the decay of an ancestor gluon
having energy fraction $x'=x/z>x$.  The loss term describes 
the disappearance of a gluon with energy fraction $x$ via the decay $x\to \{zx\,,  (1-z)x\}$, 
with $0 < z <1$. In writing the equation above, we have used the symmetry of the kernel under the exchange $z\to 1-z$. This has the consequence that the original singularity at $z=0$ has been transferred to $z=1$. As written, the loss term is divergent at $z=1$. However, it is easily checked that this divergence is cancelled by an analogous one present in the gain term (see Ref.~\cite{Blaizot:2013vha} for details).

The characteristic time 
\beq\label{stop-time}
{t_\ast}\equiv\frac{1}{\bar\alpha}\,\tau_{_{\rm br}}(E) = \frac{1}{\bar\alpha}\sqrt{\frac{E}{\hat q}},
\eeq
where $ \bar\alpha\equiv {\alpha_s N_c}/{\pi}$, with $N_c$ the number of colors, will be  referred to as the stopping time \cite{Arnold:2009ik}. It is the time at which most of the energy of the incoming parton  has been radiated into soft gluons (with typically  $x\lesssim 0.1$).  This time $t_\ast$ differs by one inverse power of the coupling constant from the typical time, $\tau_{_{\rm br}}(E)=\sqrt{E/\hat q}$, it takes the gluon of energy $E$ to branch into two gluons with momentum fractions of order one, in short the time it takes the initial parton to make one democratic branching.  Let us recall that in the BDMPS-Z theory, the branching time $\tau_{_{\rm br}}(\omega)=\sqrt{\omega/\hat q}$ of a gluon with energy $\omega$ results form the matching between the usual formation time $\tau=\omega/k_\perp^2$, and the typical tranverse momentum squared acquired during that time $\tau$ via collisions with the medium consituents, i.e., $k_\perp^2=\hat q \tau$, with $\hat q $ the jet quenching parameter. 

The  parameter $\hat q$, called the jet-quenching parameter, is a characteristic transport coefficient that enters the BDMPS-Z mechanism. It controls the momentum broadening (the average momentum squared acquired by a parton propagating in the medium for a time $\tau$ is $\langle k_\perp\rangle^2\sim \hat q \tau$ (as just used above), as well as the energy lost by this parton during its propagation, $\Delta E\sim \hat q \tau^2$.  It is valid for frequencies, $\omega_{\rm BH}\lesssim \omega\lesssim \omega_c$, where (parametrically) $\omega_c\sim \hat q L^2$, with $L$ the length of the medium, is the maximum energy that can be taken away by a single 
gluon (the mechanism that dominates the energy loss). When $\omega\gtrsim\omega_c$, the spectrum of medium-induced radiation drops rapidly. The lower limit is that of incoherent emissions, and is reached when the branching time is of the order of the mean free path between successive collisions. For energies $\omega\lesssim\omega_{\rm BH}$, the suppression due to coherence between successive emissions (the so-called Landau-Pomeranchuck-Migdal effect) disappears.  In this paper, we shall ignore this lower bound, and assume that the medium induced gluon splittings is described by the BDMPS spectrum all the way down to zero energy, and refer to the corresponding cascade as to the {\em ideal} BDMPS cascade.

In writing Eq.~(\ref{Dfin2}), we have isolated the dependence of the BDMPS kernel on the energy fraction of the parent gluon, the factors  $1/\sqrt{x}$ in Eq.~(\ref{Dfin2}), from the $z$ dependence of the splitting process. Thus, the  kernel ${\cal K}(z)$  in Eq.~(\ref{Dfin2}) is a ``reduced'' kernel  given by
\beq\label{hatK}
{\cal K}(z)=\,\frac{[1-z(1-z)]^{5/2}}
{[z(1-z)]^{3/2}}=\frac{[f(z)]^{5/2}}
{[z(1-z)]^{3/2}}.
\eeq
This kernel collects contributions from  the $z$ dependence of the actual branching time (left out in the definition\footnote{The actual branching time for offsprings carrying fractions $z$ and $1-z$ of the initial energy is $\sqrt{xz(1-z)}t_\ast$.} of $t_\ast$, Eq.~(\ref{stop-time}) above), and from the leading order splitting function $P_{gg}(z)=N_c [f(z)]^2/z(1-z)$  (we restrict our discussion to purely gluonic cascades).
 The factor $\sqrt{x}$ may then be combined with $t_\ast$ in order to define an effective time scale for the branching of a gluon carrying the fraction $x$ of the initial energy,  $t_\ast(x)=t_\ast\sqrt{x}$. This time scale decreases as $x$ decreases, meaning that the rate of emission of soft gluons is higher the softer the gluon: it follows that gluon splittings occur faster and faster as one moves down the cascade, and, as a result, it takes a finite time (of order $t_\ast$) to transport a finite amount of energy form $x=1$ down to $x=0$ \cite{Arnold:2009ik}.  It is actually convenient to make this $x$-dependent time scale more evident, and write Eq.~(\ref{Dfin2}) as 
 \beq
\label{Dfin5a}
\frac{\partial}{\partial t}D(x,t)=\int_x^1 \rmd z\, {\cal K}(z)\, \frac{D\left({x}/{z},t\right)}{t_\ast(x/z)}
-\frac{D\left(x,t\right)}{t_\ast(x)} \int_0^1\rmd z \,z{\cal K}(z)= {\cal I}[D].
\eeq
This form of the equation plays an important role in the foregoing analysis. 

In the present study, we shall use a simplified version of the reduced kernel ${\cal K}(z)$ given in Eq.~(\ref{hatK}),  namely
\beq\label{gker2}
{\cal K}(z)\approx\frac{1}{z^{3/2}(1-z)^{3/2}}\,,
\eeq
that is, we shall replace the smooth function $f(z)$ in  the numerator by 1. 
The main motivation for using this simplified kernel is that one can then obtain the  solution analytically. This solution was already discussed in Ref.~\cite{Blaizot:2013hx}. Its explicit construction is presented in \ref{exactlargexc} (see Eq.~(\ref{Gsol2Ap})).  The simplified kernel (\ref{gker2}) captures properly the singular behaviors at $z\to 0$ and $z\to 1$, while the smooth function $f(z)$ plays a negligible role, as can be verified through a  comparison of the analytic solution with the numerical solution obtained with the full kernel (\ref{hatK}).  We shall in fact verify throughout this work that the main qualitative features of the cascade are  to a large extent independent  of the specific form of the (reduced) kernel.

Before we go into a discussion of the main features of this analytic solution, we wish to contrast this cascade of medium induced radiation with the cascade of gluons emitted in vacuum by an energetic (off-shell) parton. This is described by the DGLAP evolution \cite{Altarelli:1977zs}. A simplified version of the corresponding evolution equation for the inclusive one particle distribution reads 
\beq\label{DGLAP}
\frac{\del }{\del t} D(x,t) = \bar\alpha \int_x^1 \frac{\rmd z}{z(1-z)} D(x/z,t) - \bar\alpha\int_0^1 \frac{\rmd z}{1-z}  D(x,t),
\eeq 
where here the time variable is related to the virtuality $Q^2$ of the emitting parton, $t\equiv \ln Q^2/Q^2_0$, and again only gluons are taken into account. 
This equation has the  form of Eq.~(\ref{Dfin5a}) if one identifies  
\beq
\frac{1}{t_\ast(x)}=\bar\alpha,
\eeq
and  
\beq
{\cal K}(z)=\frac{1}{z(1-z)}.
\eeq
Thus the DGLAP equation differs from the BDMPS equation in two major aspects. First, the kernel ${\cal K}(z)$ is less singular near $z=0$ and $z=1$. Second, the rate of successive branchings is independent of the parent energy, i.e., it is constant along the cascade. We shall see that the latter property is what makes the major difference between the BDMPS and the DGLAP cascades. 

%%%%%%%%%%%%%%%%%%%%%%%%%%%%
\subsection{The ideal medium-induced QCD cascade}
%%%%%%%%%%%%%%%%%%%%%%%%%%%%%
We now return to the ideal medium-induced QCD cascade, and discuss the main features of the solution to Eq.~(\ref{Dfin2}) obtained with the simplified kernel (\ref{gker2}). 
This solution reads (see Appendix A, Eq.~(\ref{Gsol2Ap}))
\beq\label{Gsol2}
D(x,\tau)=\frac{\tau}{\sqrt{x}\,(1-x)^{3/2}}\, \exp\left(-\pi\frac{\tau^2}{1-x}\right),\qquad \tau\equiv \frac{t}{t_\ast}.
\eeq

This solution exhibits two remarkable features: a peak near $x=1$ associated with the leading particle, and a scaling behavior in $1/\sqrt{x}$ at small $x$ where the $x$ dependence factorizes from the time dependence, i.e. 
\beq\label{Gsol2-soft}
D(x,\tau)\approx \frac{\tau}{\sqrt{x}}\, \rme^{-\pi \tau^2 }.
\eeq

\begin{figure}[ht]

          \includegraphics[width=6.5cm]{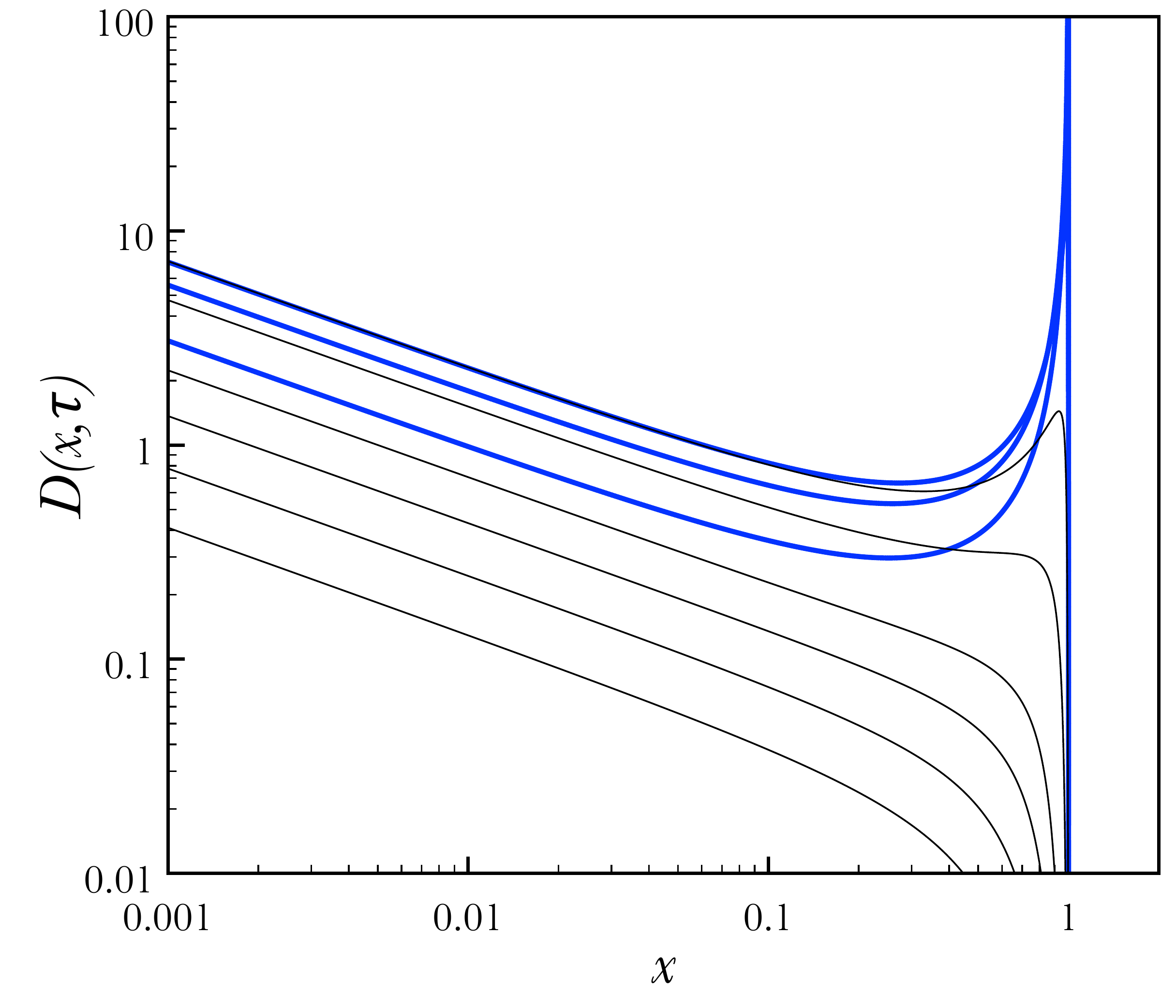} \includegraphics[width=6.5cm]{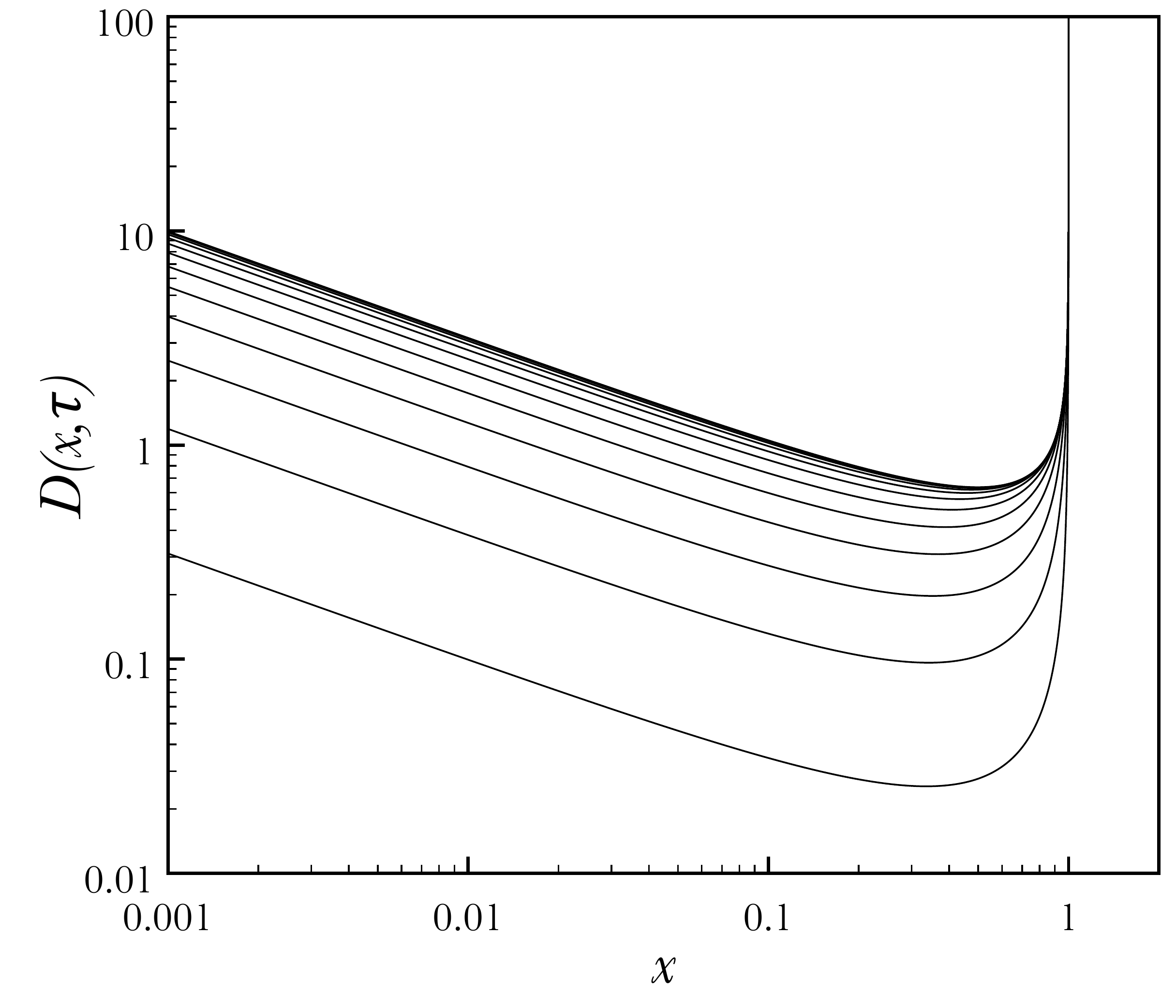} 
		\caption{(Color online.) The function $D(x,\tau)$ (Eq. (\ref{Gsol2}))  at
		various times. Left panel: the filling of the modes, which proceeds till the disappearance of the leading particle peak. 
		The values of $\tau$ are, for the thick (blue) curves, from bottom to top:  0.1, 0.2, 0.3 (during this stage the leading particle acts as a source for soft gluon radiation), and for the thin (black) curves, from top to bottom: 0.5, 0.7, 0.9,1.0,1.1,1.2  (the leading parton has exhausted its energy  and the peak has disappeared, while energy continues to flow to small $x$, the amount of energy in each mode decreasing exponentially fast). Right panel: energy is constantly injected into the system by a source located at $x=1$ (see Eq.~(\ref{DfinSource})). After a transitory regime, characterized by a uniform increase with time of the scaling spectrum, the system reaches a steady state. The values of $\tau$ are, from bottom to top: 0.1,0.2,0.3, 0.4, 0.5, 0.6,0.7,0.8,0.9,1.0. 
		 }
		\label{fig2}

\end{figure}

 An illustration of this solution is given in Fig.~\ref{fig2}, left panel. The energy of the leading particle, initially concentrated in the peak at $x\lesssim 1$, gradually disappears into radiated soft gluons,  and after a time $t\sim t_\ast$ (i.e. $\tau\sim 1/\sqrt{\pi}\approx 0.5$)  most of the energy is to be found in the form of radiated soft ($x\lesssim 0.1$) gluons. This is also the time at which the peak corresponding to the leading particle disappears (see Fig.~\ref{fig2}). These are the reasons that motivate calling $t_\ast$ the stopping time.  At the same time the occupation of the small $x$ modes increases (linearly) with time, keeping the characteristic form of the scaling spectrum. When the peak has disappeared, the cascade continues to lower $x$, causing a uniform, shape conserving, decrease of the occupations, and a flow of energy towards small $x$ that we now analyze. 

 Energy conservation is explicitly implemented in Eq.~(\ref{Dfin2}) at the level of individual splittings. To see how this conservation law manifests itself more globally, it is useful to consider the flux of energy, ${\cal F}(x_0,\tau)$, towards values of $x$ smaller than a value $x_0$. This is defined by 
\beq\label{flowdef}
{\cal F}(x_0,\tau)=-\frac{\del {\cal E}(x_0,\tau)}{\del \tau}, \qquad {\cal E}(x_0,\tau)\equiv \int_{x_0}^1 \rmd x D(x,\tau), 
\eeq
where  ${\cal E}(x_0,\tau)$ is the amount of energy contained in the modes with $x>x_0$, and ${\cal F}(x_0,\tau)$ is counted positively for energy moving to values of $x<x_0$ (hence the minus sign in the definition of ${\cal F}$. 
These quantities can be calculated explicitly. We have for instance
\beq
{\cal E}(x_0,\tau)=\int_{x_0}^1 \rmd x\, D(x,\tau) ={\rm e}^{-\pi \tau^2}\,{\rm erfc}\left(  \sqrt{\frac{\pi x_0}{1-x_0}}\,\tau\right),
\eeq
with  ${\rm erfc}(x)$ the complementary error function:
\beq
{\rm erfc}(x)=\frac{2}{\sqrt{\pi}} \int_x^\infty \rmd u\, \rme^{-u^2}.
\eeq

We note that the total energy ``stored in the spectrum'', namely
\beq
\lim_{x_0\to 0} {\cal E}(x_0,\tau)={\rm e}^{-\pi \tau^2},
\eeq
 decreases with time, and accordingly there is a non vanishing flux of energy reaching $x=0$
\beq\label{flowexactwo}
{\cal F}(0,\tau)=2\pi t\, \rme^{-\pi\,\tau^2}\,.
\eeq
 It follows that the complete, energy conserving, solution involves a contribution 
 \beq\label{condensate}
 D_c(x)=n_c(\tau)\delta(x)\quad \text{with}  \quad n_c(\tau)=1-{\rm e}^{-\pi\tau^2},
 \eeq
somewhat analogous to a condensate where the radiated energy accumulates (see next section). Note that when $\tau\sim 1/\sqrt{\pi}$, corresponding to the disappearance of the leading particles into soft radiation,  about 60\% of the initial energy has flown into the condensate.

It is interesting to consider also the situation where the leading particle is replaced by a source that injects energy at a constant rate ${\cal A}$ at $x=1$. In this case, we are led to look for the solution of the following equation
\beq\label{DfinSource}
\frac{\partial}{\partial t}D(x,\tau)={\cal A}\delta(1-x) + {\cal I}[D],
\eeq
where ${\cal I}[D]$ denotes the r.h.s. of Eq.~(\ref{Dfin2}) (or Eq.~(\ref{Dfin5a})). 
The exact solution of  Eq.~(\ref{DfinSource})  with initial condition $ D(x,\tau=0)=0$ reads
  \beq\label{Dtbex}
  D_\text{st}(x,\tau)\,=\,\frac{{\cal A}}{2\pi \sqrt{x(1-x)}}\ 
  \left(1-\rme^{-\pi\frac{\tau^2}
  {1-x}}\right),
  \eeq
and  is plotted in Fig.~\ref{fig2}, right panel. As time goes on, this  solution converges to the stationary solution 
$({\cal A}/2\pi) /\sqrt{x(1-x)}$, keeping the
shape of the small $x$ spectrum, with just an overall time--dependent scaling. Remarkably, the scaling form $D(x,t)={f(t)}/{\sqrt{x}}$ is reached well before the stationary state is achieved.
When the steady state is reached, all the energy provided by the source  flows throughout the entire system towards the condensate at $x=0$, that plays the role of a sink, while the population of the various modes stays unchanged.

%%%%%%%%%%%%%%%%%%%
\subsection{The DGLAP cascade }
%%%%%%%%%%%%%%%%%%%

We now contrast these properties with those of the DGLAP cascade. 
The solution of Eq.~(\ref{DGLAP}) is obtained in \ref{DGLAPsolution}, using a Mellin transform. The initial condition reads  $\tilde D(\nu,0)=1$, with $\tilde D(\nu,0)$ the Mellin transform of $D(x,0)=\delta(1-x)$. The solution can be expressed as the following integral (see Eq.~(\ref{DGLAP-sol-Mellin}))
\beq\label{DGLAP-sol}
D(x,t)=\int_{c-i\infty}^{c+i\infty} \frac{\rmd\nu}{2\pi i}  \exp\left[-(\psi(\nu)+\gamma)\,t+\nu\ln\frac{1}{x}\right].
\eeq
\begin{figure}[h]

		\includegraphics[width=6.8cm]{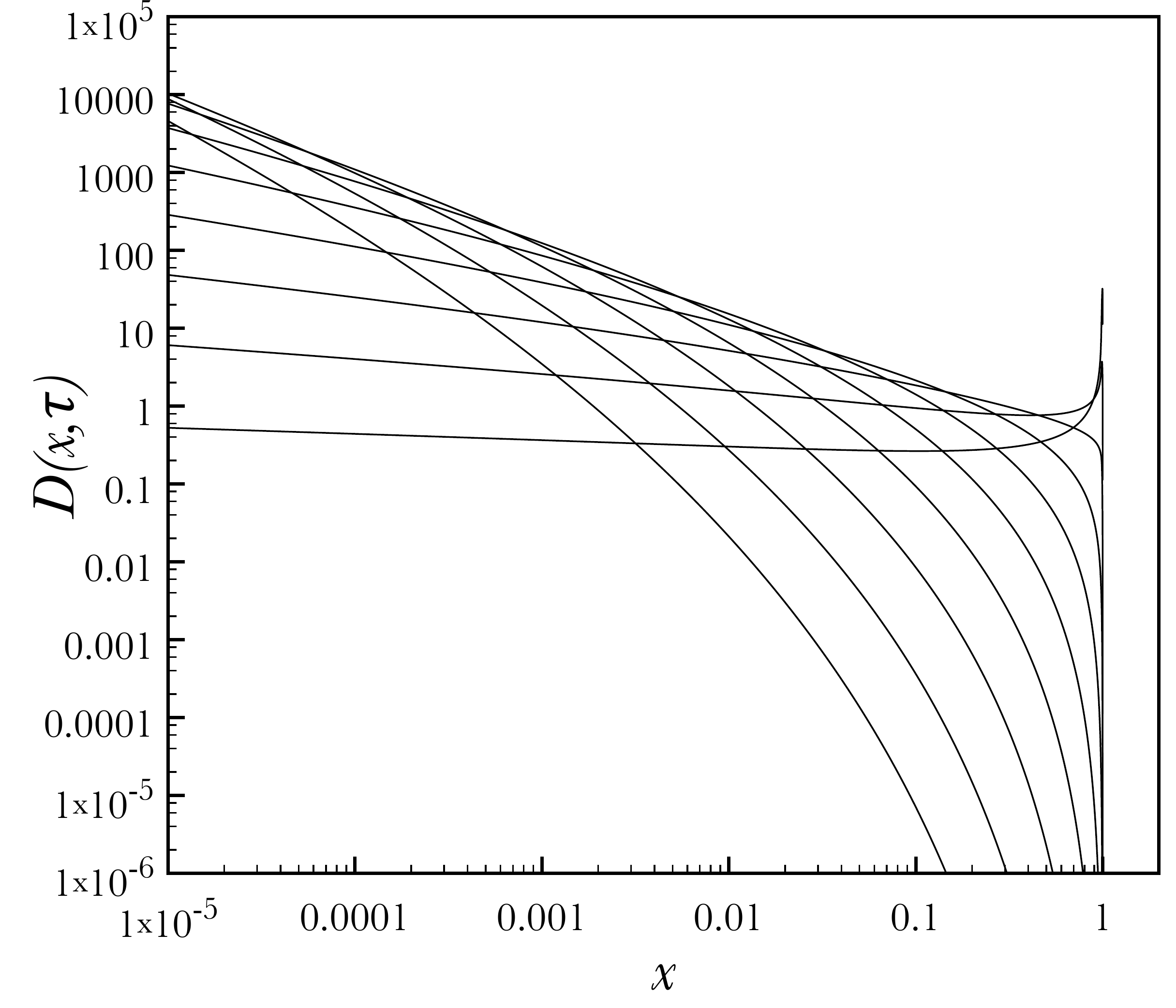} \includegraphics[width=6.8cm]{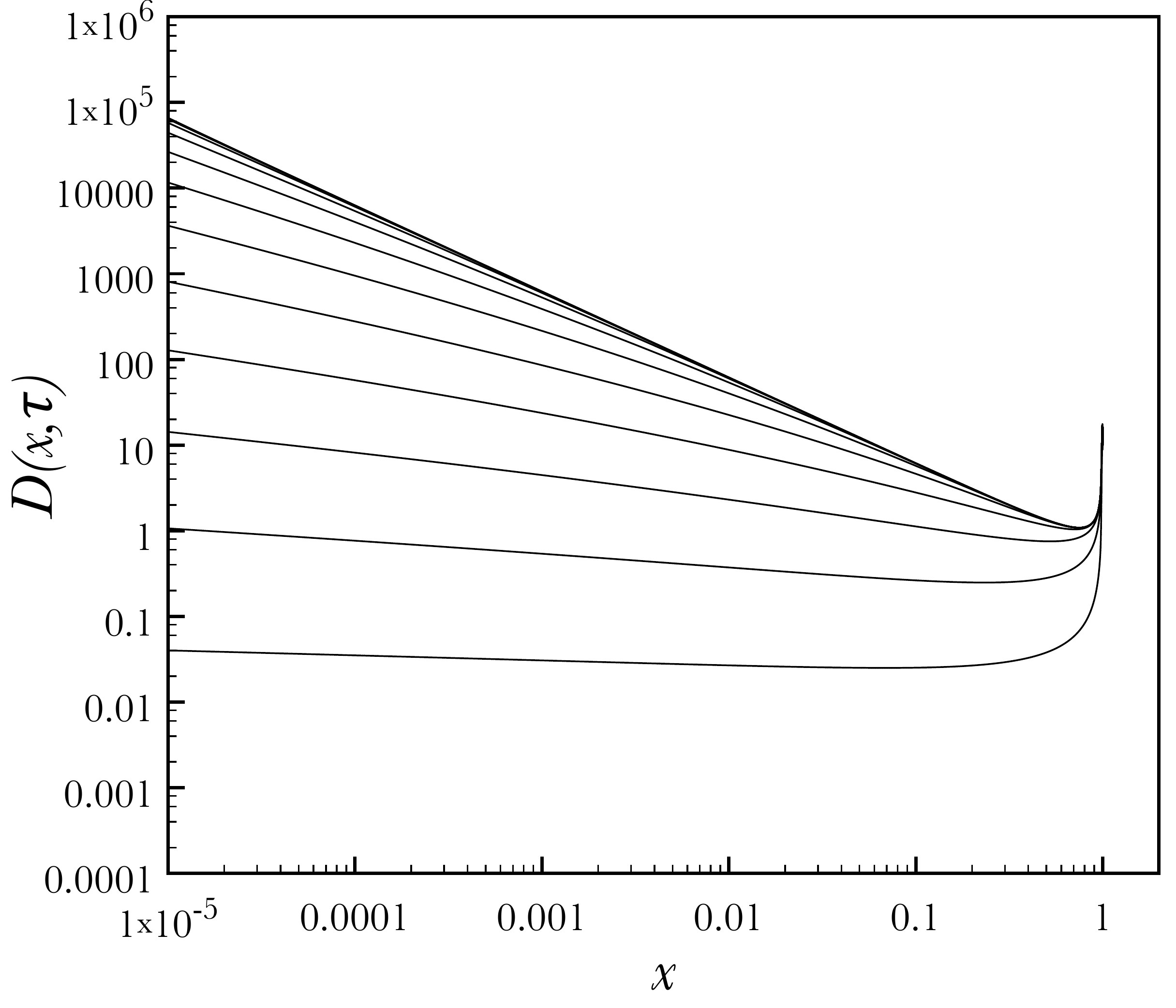} 
		\caption{Model of a DGLAP cascade according to Eq.~(\ref{DGLAP-sol}) (left panel),  and with a source added at $x=1$ (right panel). The curves corresponds to $\tau=0.2,0.4,0.6,0.8,1.0,1.2,1.4,1.6,1.8,2.0$ from bottom to top. In the right panel, the emergence of the scaling solution $D(x)\sim 1/x$ is clearly visible, as well as the persistent deviation from it of the true solution at very small $x$.}
		\label{fig:DGLAP}

\end{figure}

This integral is not easy to calculate in general. However, it allows us to verify a few properties that are relevant for our discussion. First, 
it is easy to check that  the energy is conserved by the evolution. Indeed, from Eq.~(\ref{DGLAP-sol-Mellin}) we observe that 
\beq
{\cal E}(0,t)=\int_0^1 \rmd x D(x,t) = \tilde D(1,t) =1.
\eeq
Second, the following asymptotic behavior is derived in  \ref{DGLAPsolution} (see Eq.~(\ref{DGLAPas0})): 
\beq\label{DGLAP_as}
D(x,t)\approx \left(\frac{1}{x}\right)^{2\sqrt{\frac{t}{\ln 1/x}}},\qquad\qquad (\ln 1/x \gg t).
\eeq
Here, in contrast to what happens with the BDMPS cascade, the time dependence of the energy distribution does not factorize, and no simple scaling behavior emerges at small $x$. 
 The growth of the spectrum at small $x$ is in fact tamed by the exponent in Eq.~(\ref{DGLAP_as}), making the spectrum integrable when $\ln 1/x >t $: all the energy remains in the spectrum, and no energy flows to $x=0$, unlike the BDMPS cascade.

A numerical calculation of the DGLAP cascade is presented in Fig.~\ref{fig:DGLAP}. In particular, in the right hand side panel, the model with a source is given. The solution in that case approaches the stationary solution $D(x)\sim {\cal A}/x$  at not too small $x$, but it takes a long time to populate the very small $x$ modes according to this $1/x$ law. In fact, for any time $t$, the true solution deviates from the stationary solution in the same fashion as in Eq.~(\ref{DGLAP_as}).

The properties of the two cascades that we have discussed in this section can be understood in a broader context that we now describe.

%%%%%%%%%%%%%%%%%%%%%%%%
\section{Further properties and general considerations}\label{sect:general}
%%%%%%%%%%%%%%%%%%%%%%%%

The two cascades described in the previous section are governed by the same general equation, Eq.~(\ref{Dfin5a}) with, in each case, a specific kernel ${\cal K}(z)$ and  a specific  time scale $t_\ast(x)$.  In order to get further insight into the generic features of these cascades and understand why and how they differ, it is useful to extend the discussion and regard ${\cal K}(z)$ and $t_\ast(x)$ as independent quantities. The kernel ${\cal K}(z)$ controls how, in a given splitting,  the energy is shared between the two offsprings. The time scale $t_\ast (x)$ controls the rate at which successive splittings occur, and this is allowed to depend on $x$. The independent choices made for ${\cal K}(z)$ and for $t_\ast(x)$ may appear somewhat artificial since in a given field theory the two quantities are determined by the same microscopic physics, as the examples treated explicitly in the previous section show.  This  separation, however, allows us to better illustrate the specific roles of  ${\cal K}(z)$ and of $t_\ast(x)$.

Equation~(\ref{Dfin5a}) has a number of interesting properties that we now examine. By direct integration of Eq.~(\ref{Dfin5a}), assuming that $D(x,t)$ is regular in the interval $[0,1]$,  it is easy to check that the total energy is conserved, namely, 
\beq
\frac{\del}{\del t}{\cal E}(0,t)=\frac{\del}{\del t}\int_0^1 \rmd x D(x,t)=0.
\eeq
Since energy is continuously moving towards the lower values of $x$, the distribution will generically develop a divergent behavior at small $x$. The effect of this divergence, in particular on the regularity of $D(x,t)$ as $x\to 0$,  will be discussed shortly. 

One can also verify that Eq.~(\ref{Dfin5a}) admits an approximate fixed point solution, which we refer to as a scaling solution, of the form
\beq\label{scal-spect}
D_{\rm sc}(x,t)= \frac{t_\ast(x)}{x}. 
\eeq
Indeed, in the region where $D(x,\tau)\sim D_{\rm sc}(x,\tau)$, there is complete cancellation between gain and loss terms.  Note that this compensation occurs only for $z \ge x$ in the integrals of Eq.~(\ref{Dfin5a}), so that the scaling solution cannot be an exact fixed point of the equation. We shall see however that the general solution is driven to 
 this approximate fixed point, and it approaches it  in the allowed region of small $x$.

 Depending on the behavior of the scaling spectrum (\ref{scal-spect}) when $x\to 0$, 
  the approximate fixed point solution may lead to a divergent expression for the total energy ${\cal E}(0,t)$  contained in the modes between 0 and $1$. In order for this not to happen, either the fixed point solution  is never reached at very small $x$, or  $t_\ast(x)$ is an increasing function of $x$. Both cases have already been met.  The former occurs in the DGLAP cascade for which $t_\ast(x)=\text{const.}$, while the latter is that of the BDMPS cascade for which $t_\ast(x)\sim \sqrt{x}$. To make things more precise, it is convenient to consider the energy ${\cal E}(x_0,t)$ of the modes with $x\le x_0$, with $x_0$ small but finite, and its time derivative, the flow ${\cal F}(x_0,t)$ (see Eq.~(\ref{flowdef})). A simple calculation yields 
\beq\label{flowA1}
{\cal F}(x_0,\tau)=\int_{x_0}^1\rmd z \,z {\cal K}(z)\int_{x_0}^{x_0/z} \rmd x\frac{D(x,t)}{t_\ast(x)} +\int_0^{x_0}\rmd z \,z {\cal K}(z)\int_{x_0}^1 \rmd x\,\frac{D(x,t)}{t_\ast(x)}.\nn
\eeq
When $x_0\to 0$, the second term becomes negligible (assuming that the $x$-integral is not too singular as $x_0\to0$). 
The first term has a simple interpretation: it is the integral of the loss term $D(x,\tau)/t_\ast(x)$ over all values of $x$ for which $z x<x_0$. 
If $x_0$ is small enough, and in the region where both $x_0$ and $x_0/z$ are in the scaling region so that we can replace $D(x,\tau)/t_\ast(x)$ by $1/x$, the flow is simply given by 
\beq\label{v-flow}
\int_{x_0}^1\rmd z\, z{\cal K}(z) \ln\frac{1}{z}\equiv v.
\eeq
The limit  $x_0\to 0$ exists provided $z{\cal K}(z)$ is bounded from above by $1/z$ in the limit $z\to 0$, the latter condition on the kernel\footnote{This property may be considered as a precise definition of what we call quasi-democratic branching. In this situation, the role of the splitting kernel is secondary as compared to that of the branching time in determining the qualitative properties of the cascade. These properties are then qualitatively similar to that of a cascade dominated by completely democratic branching with a kernel peaked at $z= 1/2$, or in other words by local (in energy space) interactions} insures  the quasi-locality of the branchings. For the BDMPS cascade, $v=2\pi$, while for the DGLAP cascade, $v=\pi^2/6$.  

\begin{figure}[h]
\begin{center}
		\includegraphics[width=6.5cm]{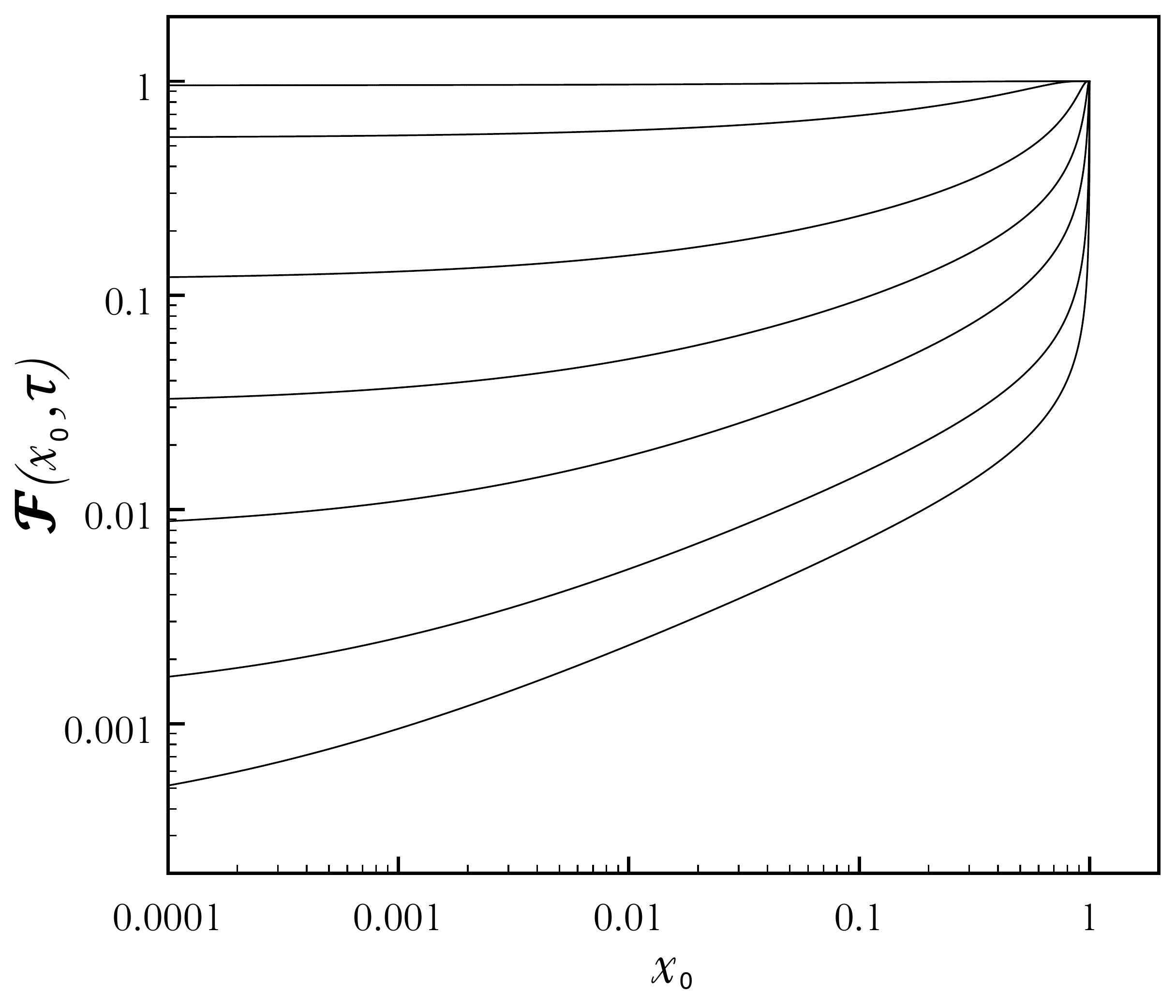} \includegraphics[width=6.5cm]{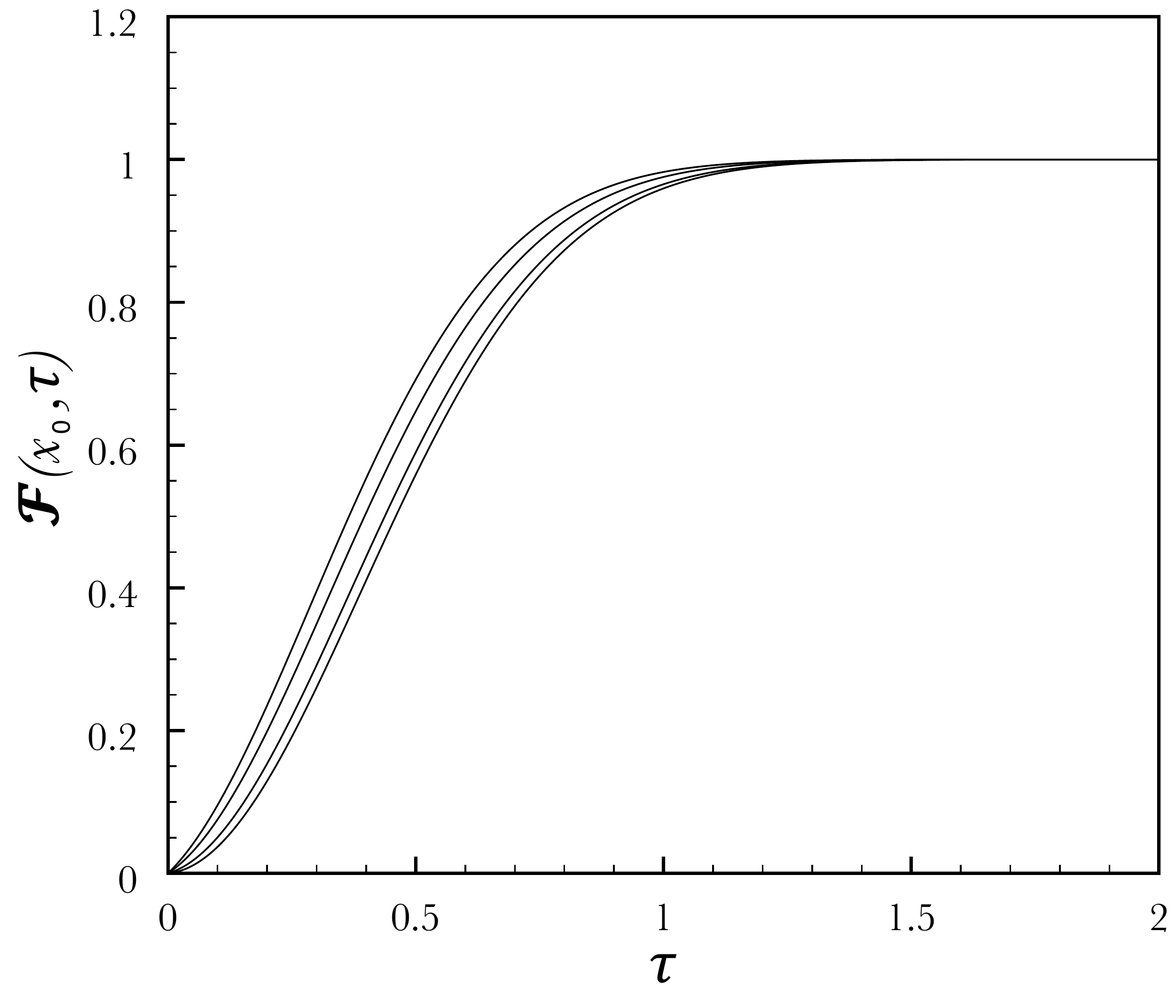} 
		\caption{ The flow in the presence of a source ${\cal A}=1$. Left:  as a function of $x_0$ for various times ($\tau=0.01, 0.02, 0.05, 0.1, 0.2, 0.5, 1$, from bottom to top). Right:  as a function of time for various values of $x_0$ ($x_0=0.1,0.05, 0.01,0.001$ from left to right). The lowest curve represents the limiting behavior of the flow at $x_0=0$, obtained from  Eq.~(\ref{Pdef2}).  }
		\label{fig:flow}
\end{center}
\end{figure}

The emergence of the scaling solution is best seen when we replace the leading particle by a source that injects energy at a constant rate ${\cal A}$ at $x=1$. That is, we look for the solution of Eq.~(\ref{DfinSource}). In this case, not only a scaling solution emerges, but it evolves into a true fixed point solution, namely a stationary solution. When a source of energy is present,  the flux of energy contains, in addition to the contribution given by the first Eq.~(\ref{flowdef}), the contribution of the source. That is
\beq
{\cal F}(x_0,\tau)={\cal A}-\frac{\del {\cal E}(x_0,\tau)}{\del \tau}=-\int_{x_0}^1 {\cal I}[D],
\eeq
where ${\cal I}[D]$ denotes the right hand side of Eq.~(\ref{Dfin5a}). 
The corresponding  flow of energy remains then given by the same formula as above, Eq.~(\ref{flowA1}). For the BDMPS cascade, an explicit calculation yields
 \beq\label{Pdef2}
 \F(x_0,\tau)={\cal A}\left[1-{\rm e}^{-\pi \tau^2} {\rm erfc}\left(\sqrt{\frac{\pi x_0}{1-x_0} }\tau  \right)     \right].
 \eeq
 This formula reproduces the result already mentioned, namely the existence of a flux of energy down to $x_0=0$.  At large times, all the injected energy flows to $x=0$, which plays the role of a sink. Eq.~(\ref{Pdef2})  also reveals how the $x_0$ dependence gradually disappears as we approach the stationary solution. This is illustrated in Fig.~\ref{fig:flow}. In the BDMPS cascade,  the scaling regime extends all the way down to $x=0$. This implies the existence of a finite flow of energy at $x=0^+$ and the formation of a condensate (or a sink) at $x=0$ where the injected energy is accumulating (see Eq.~(\ref{condensate})). 

When the stationary solution is reached, the flow is equal to that of the energy injected in the system, namely ${\cal A}$. Equating this with the general expression (\ref{flowA1}), and assuming $x_0$ to be small enough so that both $x_0$ and $x_0/z_0$ are in the scaling region, we fix the normalization of the stationary solution in the scaling region, 
\beq
D(x,\tau)\approx c \frac{t_\ast(x)}{x},\qquad c=\frac{{\cal A}}{v},
\eeq
with $v$ given by the integral (\ref{v-flow}).

For the DGLAP cascade, in  the presence of a source, one expects a steady spectrum of the form 
\beq
D_\text{st}(x) = \frac{{\cal A}t_\ast}{v\, x}.
\eeq
However this scaling  spectrum is  achieved only when $\ln 1/x \ll \ln 1/x_\text{min} \sim t$. For smaller $x$, the $1/x$ divergence is tamed, in such a way that the integral giving ${\cal E}(0,t)$ remains finite. As a result there is no flow of energy reaching $x=0$. These features are visible in the plots of Fig.~\ref{fig:DGLAP}. Thus, in DGLAP like cascades, there exists a dynamically generated infrared cutoff, $\sim x_\text{min}$, that prevents  the total energy to diverge, as it would if the scaling solution were to exist all the way down to $x=0$. At the same time, it is necessary for the integral to diverge (in the absence of this cutoff) in order to allow for an infinite amount of energy to be stored in the spectrum.  Since the finite integral  must account for the total energy that has been injected in the system after a time $t$, ${\cal A} t$, we have
\beq
{\cal A} t=\frac{{\cal A}t_\ast}{v} \int^1_{x_\text{min}} \frac{\rmd x}{x} \sim \frac{{\cal A}t_\ast}{v} \ln \frac{1}{x_\text{min}},
\eeq
so that $\ln 1/x_\text{min} \sim  t$, or $x_{\rm min}\sim \rme^{-\# t}$. Because all the injected energy can be stored in the spectrum within the interval $[x_\text{min},1]$, there is no flow at $x=0$. The cutoff $x_{\rm min}$ decreases rapidly  with time, thus allowing more and more energy to be stored in soft modes.

The discussion above leads us to expect generically two types of cascades. To keep the discussion simple, we shall assume  the following scaling form for the characteristic time: $t_\ast(x) = t_\ast x^\gamma$. The two classes of cascades correspond to specific values of $\gamma$:

\noindent{Type A: $\gamma \leq 0$ }

The integral of the scaling solution diverges when $x\to 0$. At any finite time,  the scaling spectrum does not extend to arbitrary low $x$, but only to the value $x_\text{min}$. In the presence of a steady source, the energy ${\cal E}(x_\text{min},t)$ that is stored into the system is proportional to the time $t$, 
\beq\label{xmin}
{\cal E}(x_\text{min},t)\sim t \sim \int^1_{x_\text{min}} \frac{\rmd x}{x^{1-\gamma}}\sim \begin{dcases} 
        \ln (1/x_\text{min}), \quad \text{for}  \quad \gamma=0 \\
     x_\text{min}^{\gamma} ,\quad\quad\quad~~\text{for}  \quad \gamma<0
       \end{dcases} 
\eeq
We can thus deduce the value of $x_\text{min}$  in the two cases: when $\gamma=0$, $x_\text{min}\sim \rme^{-\# t}$ and when $\gamma<0$, $x_\text{min}\sim t^{-1/|\gamma|} $. In either case the point $x=0$ is reached only in the limit $t\to\infty$.

\noindent{Type B: $\gamma > 0$ }

In this case the scaling spectrum is integrable in the infrared. This reflects the fact that it takes only a finite time to transport a finite amount of energy from $x=1$ to $x=0$, so that the energy does not accumulate in the spectrum but at $x=0$. The cascade  also involves a characteristic time, or equivalently a characteristic value $x_s $ of $x$, where multiple branchings become dominant. This is given by  $t \sim t_\ast(x_s)\sim t_\ast x^\gamma$, so that
\beq\label{xs}
x_s= \left(\frac{t}{t_\ast}\right)^{1/\gamma}.
\eeq
When $x \gg x_s$, only few branchings occur and the spectrum is well approximated by the lowest order in a perturbative expansion,
\beq
D(x,t)\sim D_1(x,t)\sim \frac{t}{t_\ast} x{\cal K}(x).
\eeq
It is in the multiple-branching regime, i.e., for   $x\lesssim x_s$, that the scaling spectrum develops,
\beq\label{xsgamma}
D_\text{sc}(x,t)= \frac{t_\ast(x)}{x} f(t),
\eeq
where $f(t)$ is a function of time that depends on the kernel ${\cal K}(z)$. Note that Eq.~(\ref{xsgamma}) indicates that $x_s\to 1$ as $\gamma\to\infty$, reflecting the fact that the scaling regime develops on the entire available energy range. 

At this point, it is useful to  comment briefly on a peculiarity of the BDMPS cascade, that makes the transition between the dilute, single-branching regime, and the the multiple branching regime invisible in the spectrum (\ref{Gsol2}). Indeed, for $x_s\ll1 $ (or equivalently $\tau\ll 1$), the spectrum is of the form
\beq
D(x,\tau)\simeq \frac{\tau}{\sqrt{x}},
\eeq  
and this is so independently of whether $x  \gg x_s$ or $x \ll x$. This property is specific to the BDMPS spectrum for which $x {\cal K}(x)\sim 1/\sqrt{x}$, so that the leading perturbative result has already the scaling form. 

%%%%%%%%%%%%%%%%%%%%%%%%%%%%%
\section{Simple models}
%%%%%%%%%%%%%%%%%%%%%%%%%%%%%

In order to illustrate the above general considerations,  we consider in this section simple models that differ by the form  of the branching time $t_\ast(x)$ and by that  of the kernel ${\cal K}(z)$.
We assume, as we did at the end of the last section,  that  $t_\ast(x)$ scales as  
$t_\ast(x)\equiv t_\ast x^\gamma$, with $\gamma$ an arbitrary parameter in the physically interesting range $0\le \gamma<1$. We shall consider two models that illustrate the generic types of cascades that we have identified in the previous discussion. In the first model, to be referred to as {\em model A}, $\gamma=0$, corresponding to a constant branching rate, as in the DGLAP cascade.  In the other model, to be referred to as {\em model B}, $\gamma=1/2$, the branching rate grows as in the BDMPS cascade.  This difference in the branching rate, coded here in the value of  $\gamma$, has a profound impact on the solution: if the branching time is independent of $x$,  it takes an infinite time to populate the smallest $x$ modes. In the case where $\gamma>0$, the splitting rate increases as we move down in $x$, and it takes only a finite time to transfer energy from the highest to the lowest values of $x$.  To emphasize that the $x$-dependence of the branching rate is the dominant factor that determines the main features of the cascade, we shall consider two extreme versions of splitting kernel. The first corresponds to  democratic branching, with ${\cal K}(z)$ peaked at the value $1/2$. The second is a flat kernel, independent of $z$. We start by analyzing models with democratic branching.\\

%%%%%%%%%%%%%%%%%%%%
\subsection{Democratic branching}
%%%%%%%%%%%%%%%%%%%%
 
Complete democratic branching corresponds to a kernel of the form
\beq
{\cal K}(z)=2\delta (z-1/2),\qquad \int_0^1\rmd z \,z{\cal K}(z)=1.
\eeq
With this choice of ${\cal K}$, Eq.~(\ref{Dfin5a}) becomes
 \beq\label{Dfin5b}
\frac{\partial}{\partial t}D(x,t)=2\theta(1/2-x)\,\frac{D\left({2x},t\right)}{t_\ast(2x)}-\frac{D\left(x,t\right)}{t_\ast(x)},
\eeq
where the theta function in front of the gain term reminds us that the support of $D(x,t)$ is limited to $x<1$. This theta function will often be omitted in the following (the constraint on the support of $D$ being implicitly implemented in the function $D$ itself). 

The two models that we consider correspond to different choices for $t_\ast(x)$. For model A, $t_\ast(x)=t_\ast$ is constant, and Eq.~(\ref{Dfin5b}) reads
\beq\label{kin-A}
\del_\tau D(x,\tau)=2D(2x,\tau)-D(x,\tau),\qquad \mbox{        \rm (model A)}
\eeq
with $\tau\equiv t/t_\ast$. It admits the following, non integrable, fixed point solution, 
$D(x)=\frac{1}{2x}$. For model B,  $t_\ast(x)=t_\ast \sqrt{x}$, and Eq.~(\ref{Dfin5b}) reads
\beq\label{kin-B}
\del_\tau D(x,\tau)=\frac{1}{\sqrt{x}}\left[ \sqrt{2}D(2x,\tau)-D(x,\tau)\right],\qquad \mbox{        \rm (model B)}
\eeq
with again $\tau\equiv t/t_\ast$. It admits the fixed point solution $ D(x)=\frac{1}{\sqrt{x}}$.
In contrast to model A, the singularity at small $x$ is now integrable.

When the energy is  injected in the system at exactly $x=1$ (either by a leading particle or by a source),  the distribution function has support on a discrete set of points. It is of the  form 
\beq
D(x,\tau)=\sum_{k=0}^\infty \,C_k(\tau)\,\delta\left(\frac{1}{2^k}-x  \right). 
\eeq
The equation (\ref{kin-A}) then leads, for both models, to the following recursion relations for the coefficients $C_k(\tau)$ 
\beq\label{recusrionCk}
\dot C_0={\cal A}-C_0,\qquad \dot C_k=C_{k-1}a^{k-1}-C_ka^{k},
\eeq
with\footnote{In this section, we focus on the case $\gamma=1/2$. A solution for general values of $a$ is presented in \ref{exactdemo}.}
\beq
{\mbox  {\rm model A:}   }\;\;\; a=1,\qquad\qquad\qquad {\mbox  {\rm model B: }  } \;\;\;a=2^\gamma,
\eeq
and ${\cal A}$ the strength of the external source. When ${\cal A}$  is non zero, the initial condition is $C_k=0$ for all $k$. When ${\cal A}=0$, the initial condition is $C_0=1$ and $C_k=0$ for all $k\ne 0$. In either case, the equation for $C_0(\tau)$ is the same for both models, and it  can be solved easily:
\beq
{\cal A}=0:\quad C_0(\tau)=\rme^{-\tau},\qquad   {\cal A}\ne 0:\quad C_0(\tau)={\cal A}(1-\rme^{-\tau}).
\eeq 
When ${\cal A}=0$, the exponential decay of $C_0(\tau)$ represents the decay of the leading particle. When ${\cal A}\ne 0$, this decay competes with the feeding of the zeroth mode by the source. 
The fixed point solution corresponds to $C_{k-1}=a C_k$. Whether it is reached or not depends on the initial condition, and on $\gamma$. 
The flow at $x_0=1/2^{k_0}$ is given by
 \beq\label{flowA2}
{\cal F}(k_0,\tau)={\cal A}-\sum_k^{k_0}\dot C_k(\tau)=a^{k_0}C_{k_0}(\tau).
\eeq
This equation has a simple interpretation: the flow of energy that moves to modes with values of $k$ higher than $k_0$ is equal to the loss term, $a^{k_0}C_{k_0}$,  of the equation (\ref{recusrionCk}) for the coefficient $C_{k_0}(\tau)$.\\

 We now analyze successively the solutions of models A and B. \\
 
\subsubsection{\it Model A}

\begin{figure}[ht]

		\includegraphics[width=6.5cm]{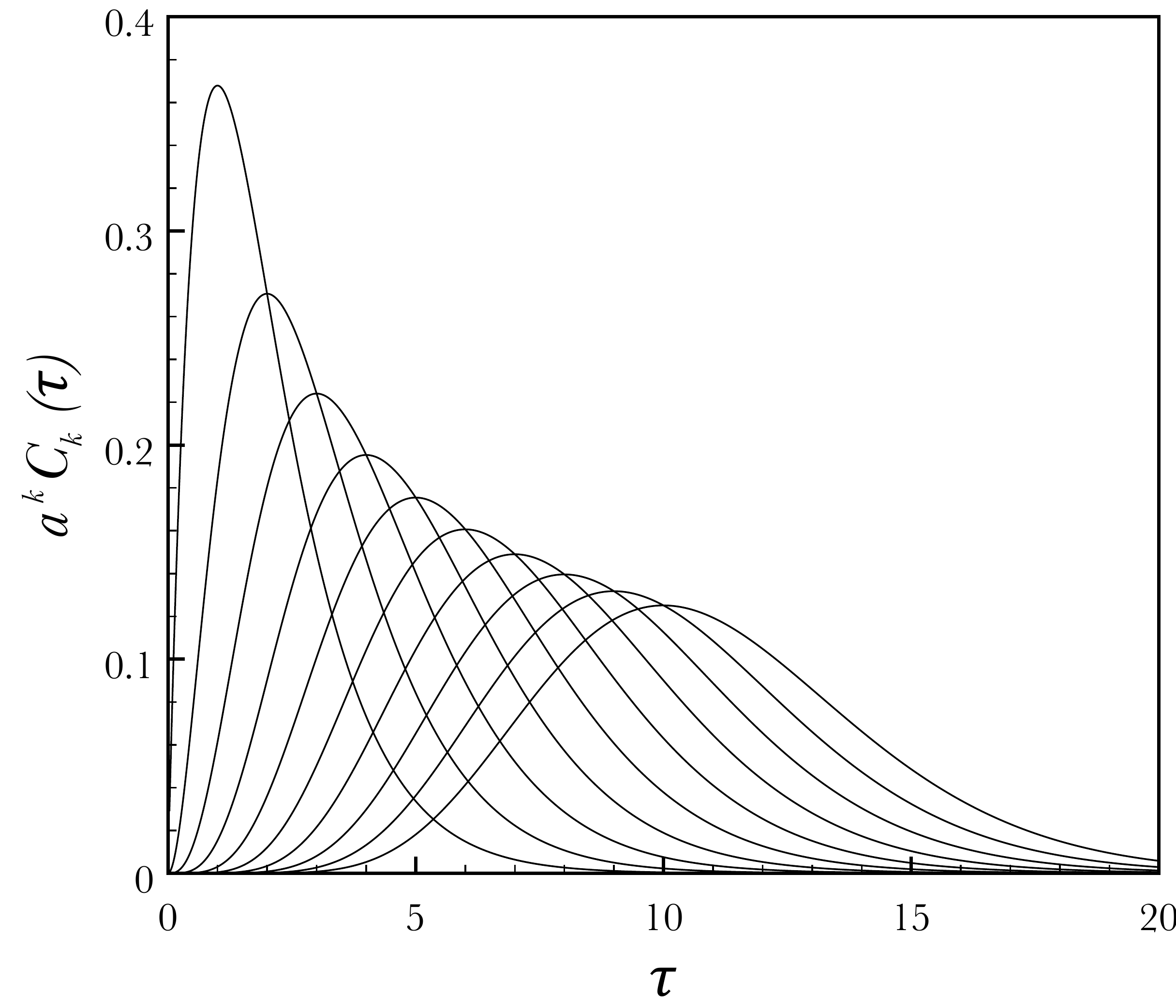}\includegraphics[width=6.5cm]{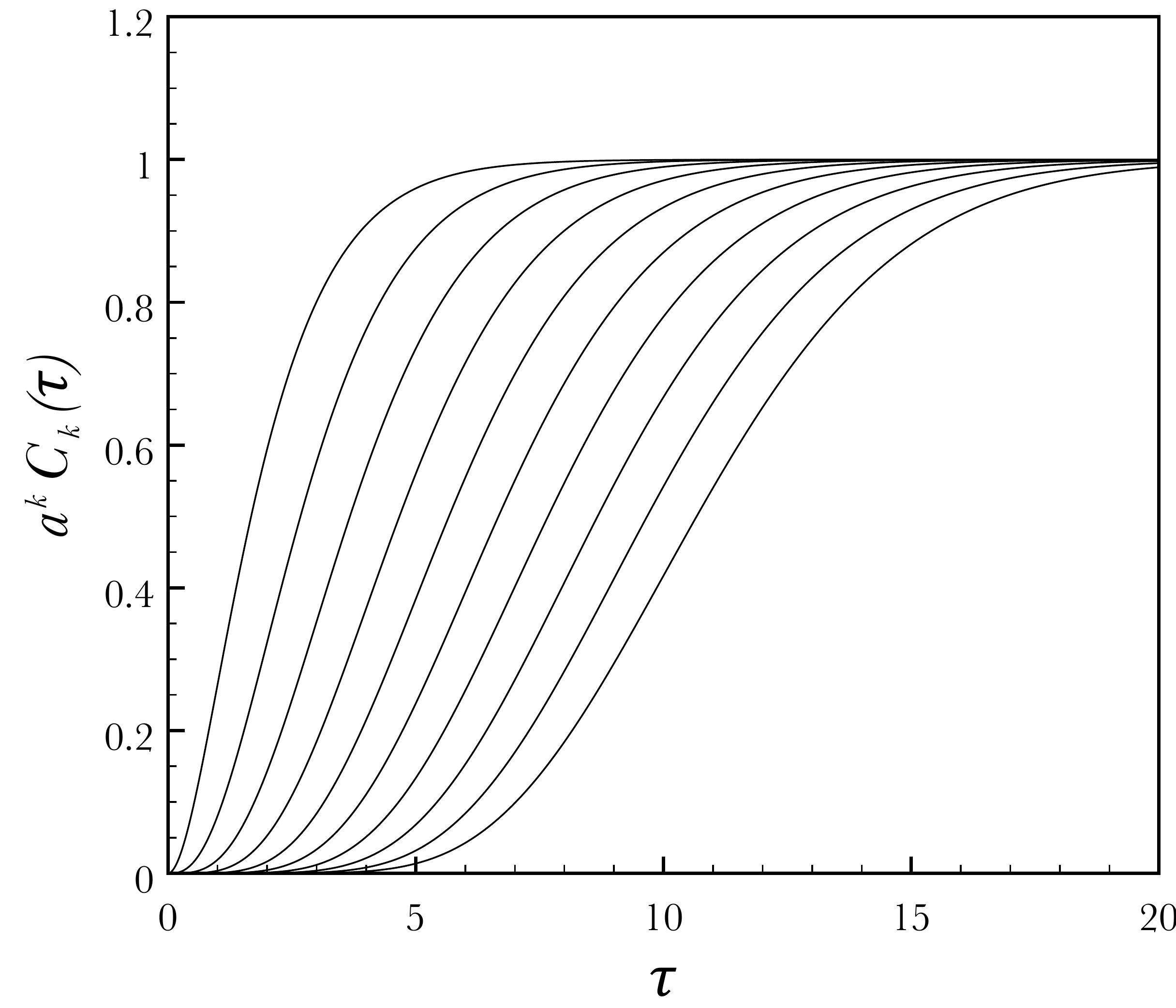} 
		 \includegraphics[width=6.5cm]{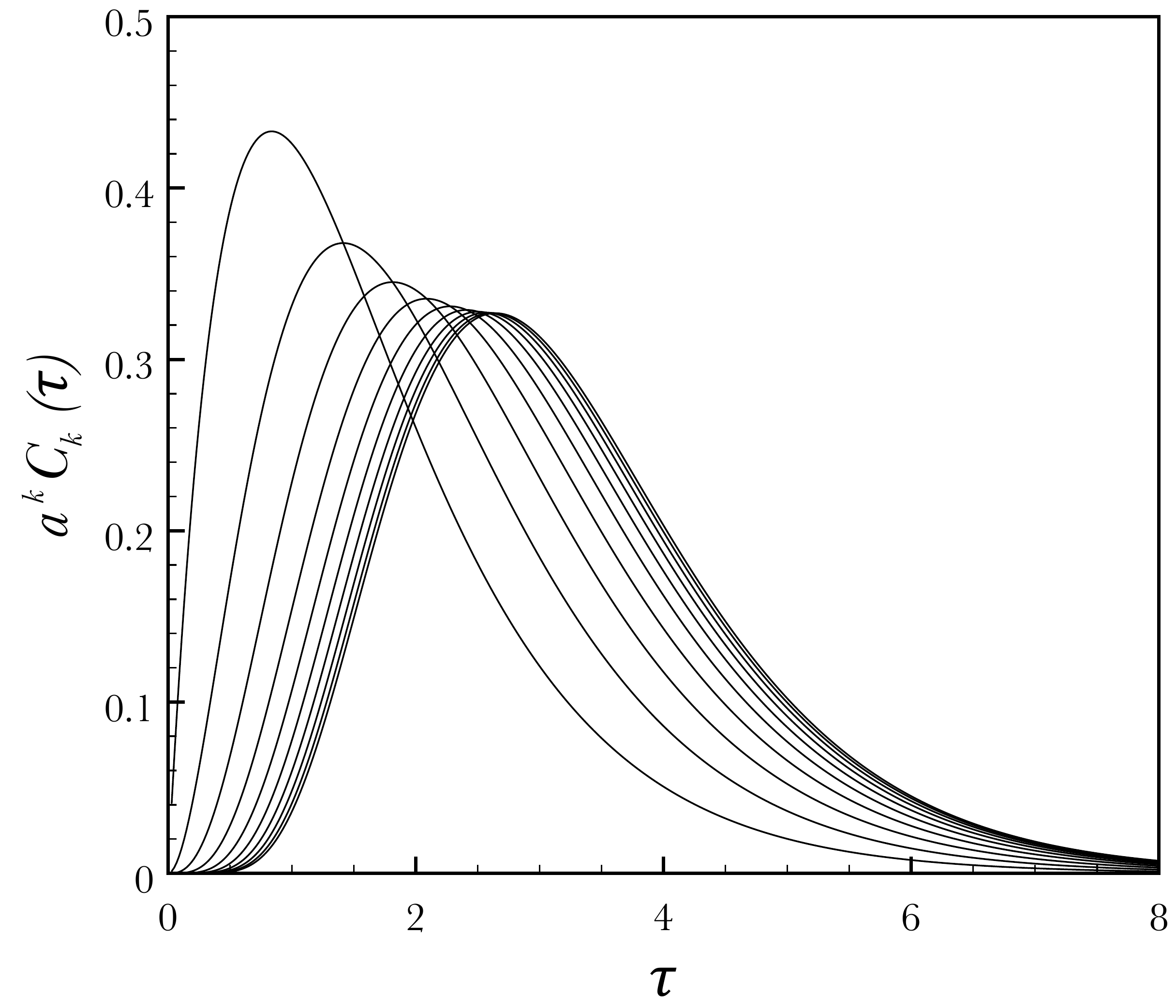}\includegraphics[width=6.5cm]{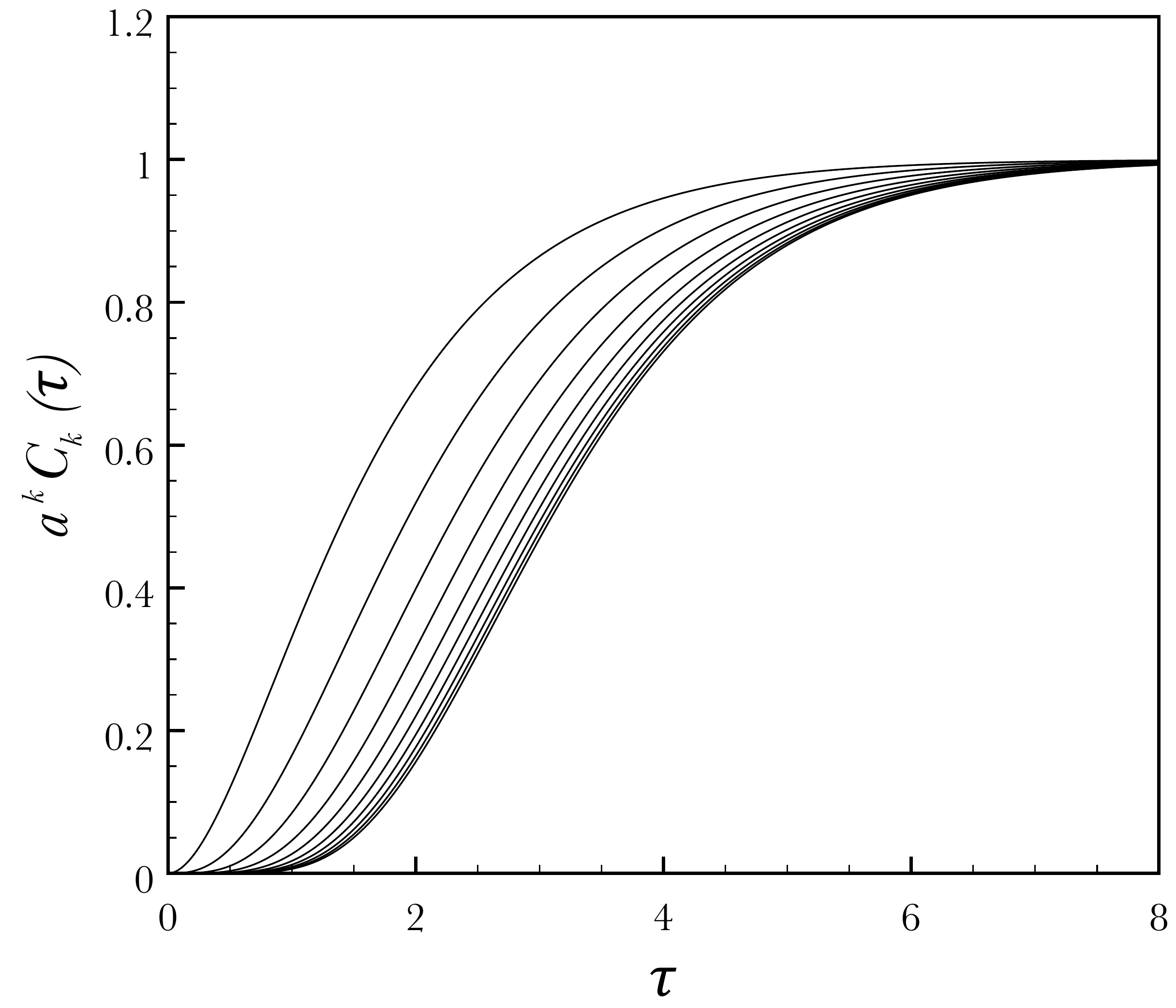} 
		\caption{The coefficients $C_k$ for the solution with a leading particle (left panels) and with a source (right panels), for the following values of $k$, $1,2,3,4,5,6,7,8,9$, from left to right. The top panels correspond to model A, the lowest panels to model B.}
		\label{figb1}

\end{figure}

 We first examine the solution  corresponding to a leading particle carrying initially all the energy. The initial condition in that case is $D(x,\tau=0)=\delta(1-x)$.
The coefficients $C_k$  obey a Poisson law,  
\beq\label{Ck-A}
C_k(\tau)=\frac{1}{ k!} \tau^k{\rm e}^{-\tau}.
\eeq
 The energy of the leading particle ($k=0$) decreases exponentially with time, $C_{k=0}=\rme ^{-\tau}$, as already noticed. The energy of the mode $k_0$ goes through a maximum: it increases till $\tau=k_0$ as it is filled by energy coming from the modes with $k<k_0$, and then it decreases exponentially when the energy of these feeding modes is exhausted.  It takes therefore an infinite amount of time to populate the modes with $k_0\to\infty$. The total energy is conserved, $\int_0^1\rmd x D(x,\tau)=1=\sum_{k=0}^\infty C_k(\tau)$: the very small $x$ (large $k$)  modes are in fact never populated. Near $x=0$ there is no flow, since for any given $\tau$, $C_k(\tau)\to 0$ as $k\to \infty$. Note also that it takes a finite time, $\tau\sim k$ to excite modes with a given $k$: this is reflected in the fact that in Eq.~(\ref{Ck-A}) giving the coefficient $C_k(\tau)$, the first $k$ coefficients of the Taylor expansion in powers of time vanish. All these features can be recognized in Fig.~\ref{figb1}, upper left panel.

Consider then the case with a source ${\cal A}\ne 0$. 
The $C_k$'s are then given by 
\beq\label{CkAws}
C_k(\tau)={\cal A}\left[ 1-\rme ^{-\tau}\sum_{n=0}^k\frac{\tau^n}{n!}  \right].
\eeq
 They become eventually all equal to ${\cal A}$ as $\tau\to\infty$. However reaching this fixed point takes an infinite amount of time. Indeed,  the function of $\tau$ that multiplies ${\cal A}$ is a growing function of $\tau$ that saturates at 1 for large $\tau$, the time at which saturation is  reached being a growing function of $k$:  the small $k$ modes are  rapidly filled, but it takes a  time of order $k$  to fill the modes with large $k$.  The flow at $x_0=1/2^{k_0}$ is equal to $a^{k_0}C_{k_0}(\tau)=C_{k_0}(\tau)$ and, for any finite $\tau$, it vanishes   for $k_0\gg\tau$, because the $k_0$ modes have not have time to get excited. These features are illustrated in Fig.~\ref{figb1}, upper right panel.

\subsubsection{Model B}

The solutions  for model B, with or without a source,  are obtained in \ref{exactdemo}. They are plotted in Fig.~\ref{figb1}, lower panels. The striking difference with model A resides in the fact that  these solutions quickly approach a scaling regime, where the time dependence factorizes and the shape of the spectrum gets frozen: it becomes independent of $k$ for $k$ large enough.  This is intimately connected to the fact that, in model B,  it takes a finite amount of time to transport energy from $x=1$ to $x=0$, in contrast to model A where this time is infinite. 

This difference between the two situations is best illustrated by considering an even simpler version of the previous models, in which the splittings occur at discrete time steps.  After $n$ time steps, the energy of the leading particles is shared equally by $2^n$ particles. It takes an infinite number of time steps to transfer all the energy of the leading particle into an infinite number of particles carrying a vanishing amount of energy. However, whether this infinite number of time steps takes an infinite amount of time depends on the model considered.  In model A, we  assume that the time interval between two splittings is a constant $\Delta\tau$. In this case, it takes an infinite time $T=\sum_{n=0}^\infty \Delta \tau$ to reach the end of the cascade. In model B, we assume that the time between two successive  branchings decreases as one moves down the cascade, according to the relation
\beq
\Delta\tau=\Delta \tau_0\left(\frac{1}{a}\right)^n,\qquad a=2^\gamma,
\eeq
where $\Delta \tau_0$ is the time it takes to make the first branching.
The time $T_N$ it takes to go through $N$ successive branchings is given by \cite{Arnold:2009ik}
\beq
T_N=\Delta\tau_0\sum_{n=0}^N \left(\frac{1}{a}\right)^n=\Delta\tau_0\frac{1-a^{-N}}{1-a^{-1}}.
\eeq
For model A, $T_N=N\Delta \tau_0$ diverges as $N\to\infty$. However, for $\gamma=1/2$, $\lim_{N\to\infty} T_N$ is finite, and $T_N$ converges in fact rapidly towards its limit. For instance, for $a=\sqrt{2}$, $1-a^{-N}>0.9$ already for $N=7$. This number $N=7$ is roughly the number of curves that one can distinguish in Fig.~\ref{figb1}, lower panels. Thus in this model B, the entire energy of the leading particle is transferred, in a finite time, to an infinite collection of particles carrying each  a vanishing amount of energy. This is the analog, in this discrete model, of the condensate that we discussed earlier: the energy injected in the system eventually ends up at $x=0$ in a finite time. \\

The fact that energy is transferred to $x=0$ in a finite time also helps us to understand the factorization of the time dependence and the emergence of the scaling solution. 
The plots in Fig.~\ref{figb1}  indicate clearly that the filling of the modes takes place in an essentially finite time, after which the solution acquires a (nearly) factorized form, with the time dependence of $a^k C_k(\tau)$ becoming independent of $k$, that is $C_k\sim 1/a^k$.  In the case where energy is injected into the system by a source ${\cal A}\ne0$, a stationary regime is eventually reached, but the scaling solution emerges well before that, as the plot in Fig.~\ref{figb1} suggests.

\begin{figure}[h]

		\includegraphics[width=6.5cm]{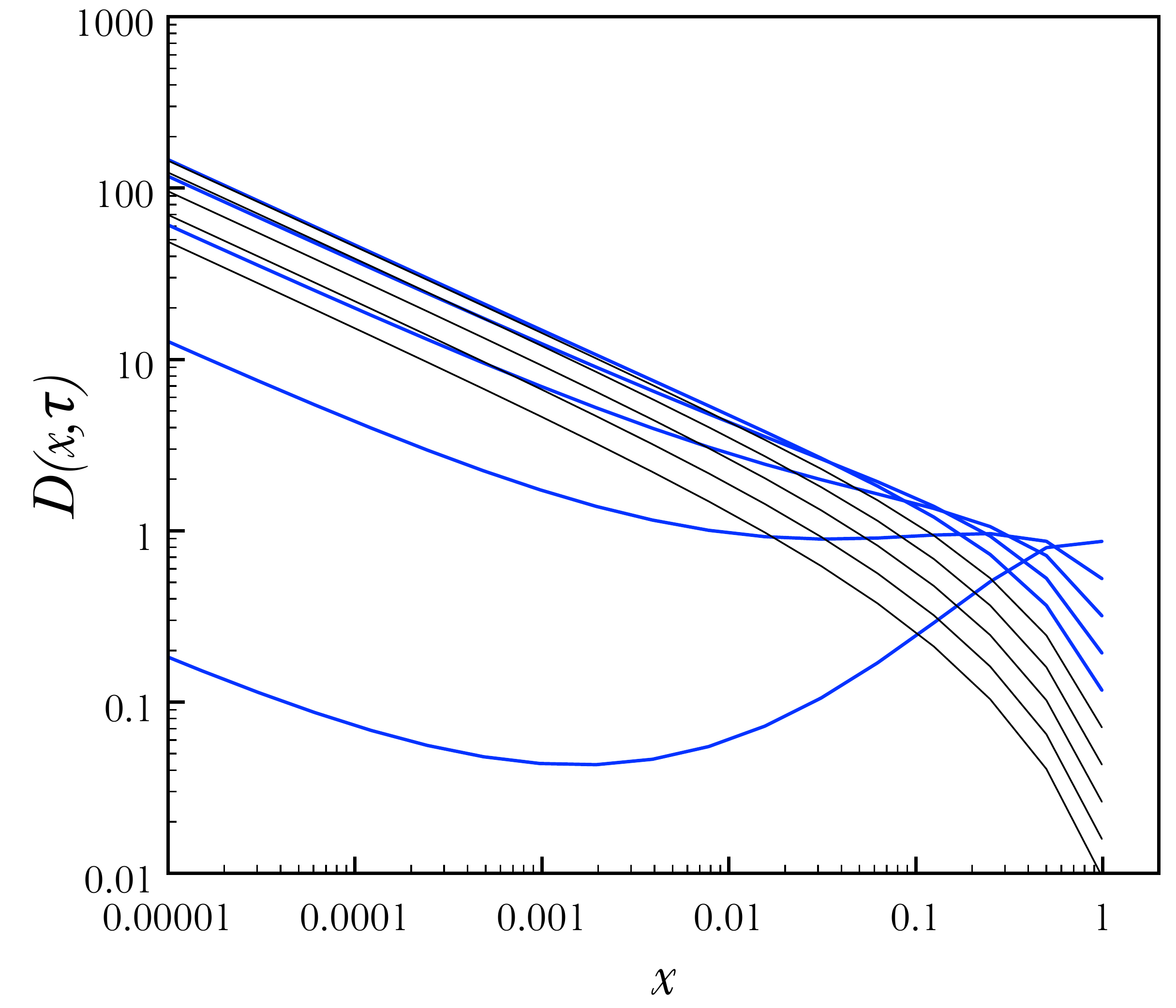} \includegraphics[width=6.5cm]{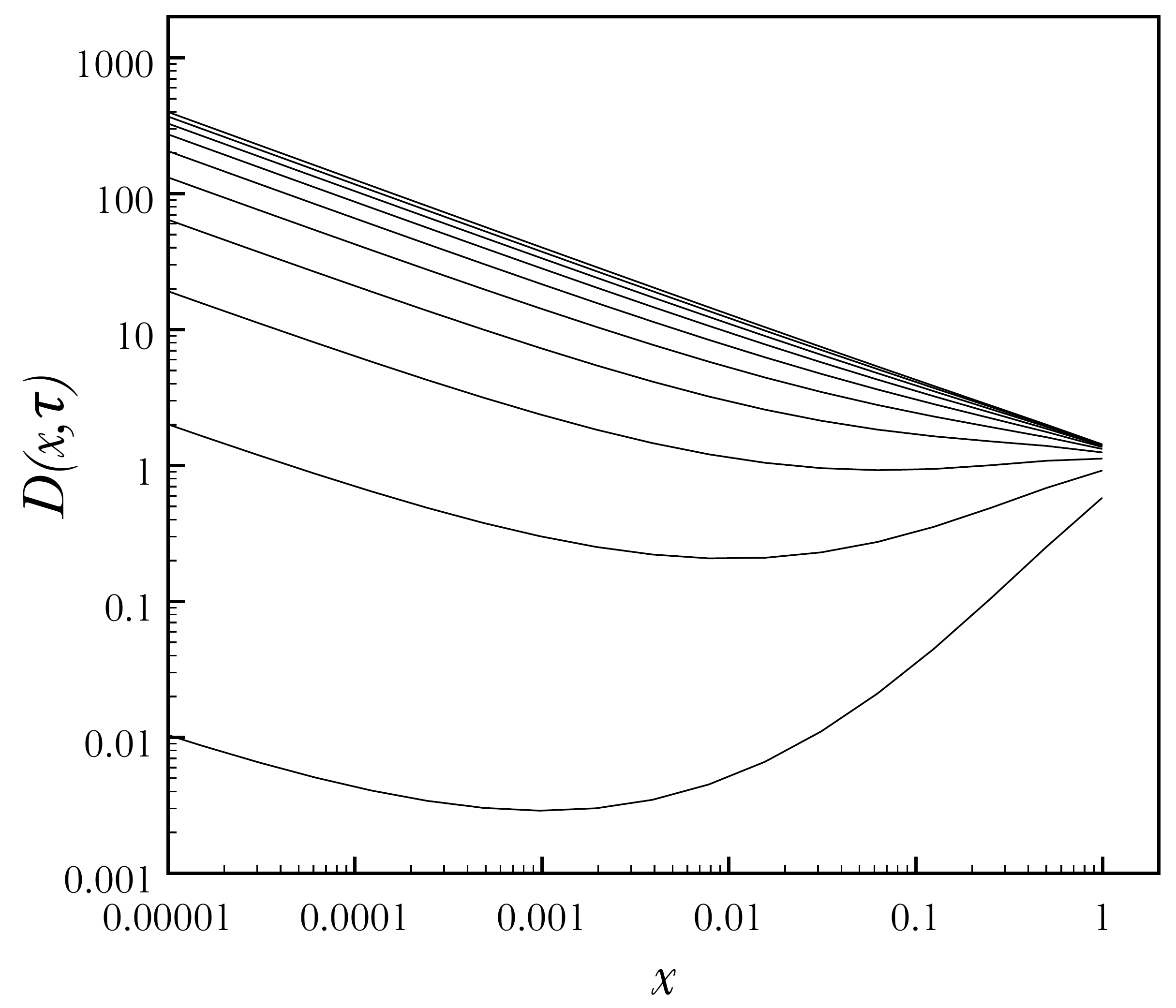} 
		\caption{(color online.) The emergence of the scaling  behavior for the distribution $D(x,\tau)\sim C_k /x $ as a function of $x=1/2^k$  for $a=\sqrt{2}$ (the discrete set of points on which the function is defined have been joined by a continuous curve). Left, the leading particle case: thick (blue) curves are for $\tau=0.5,1,1.5,2$ from bottom to top. Thin (black) curves are for $\tau=2.5,3,3.5,4.5,5,5.5$ from top to bottom. One sees the disappearance of the leading particle and the uniform growth with time of the small $x$ (large $k$) modes. When the leading particle that was feeding energy into the small $x$ modes has disappeared, the energy continues to flow to $x=0$, the populations of the various small $x$ modes decreasing uniformly with time. Right panel: Energy is injected in the system at a constant rate by a source located at $x=1$. The various curves correspond to $\tau=0.5,1,1.5,2,2.5,3,3.5,4.5,5,5.5$, from bottom to top. }
		\label{scalingA}

\end{figure}

The scaling behavior observed in Fig.~\ref{figb1} implies that the dominant contribution of the  coefficients $C_k$ given by Eq.~(\ref{CkBwos}) tends to a factorized form 
\beq\label{scalingck}
 \frac{1}{a^k}\, f(\tau).
\eeq
This can indeed be inferred from Eq.~(\ref{CkBwos}). 
 At late times, the function $f(\tau)$ is given by the term proportional to $\rme^{-\tau}$:
\beq
f(\tau)\simeq \rme^{-\tau} \lim_{k\to\infty}\prod_{m=1}^k\frac{a^{m}}{a^m-1}=\rme^{-\tau}\prod_{m=1}^\infty\frac{a^{m}}{a^m-1},
\eeq
the other terms in the series of exponentials $e^{-a^m t}$ becoming rapidly negligible. One can in fact relate this scaling behavior to that expected for the corresponding continuum solution of Eq.~(\ref{kin-B}). 
To do so, let us focus on the small $x$ region where the density is large enough to allow us to replace the sum over $k$ by an integral
\beq
\sum_k \simeq \int \rmd k \equiv \frac{1}{\ln 2}\int \frac{\rmd x}{x}, 
\eeq
with $x=1/2^k$. We have then
\beq\label{scalingckcont}
D(x,\tau)&=&\sum_{k=0}^\infty C_k(\tau) \delta(x-2^{-k})\nn
&\approx &\frac{1}{\ln 2}\int \frac{\rmd x'}{x'} C_{k(x')}(\tau) \delta(x-x')\nn
&= &\frac{1}{x \ln 2 }C_{k(x)}(\tau).
\eeq
Now, using the fact that $a=2^\gamma$, we rewrite Eq.~(\ref{scalingck}) as 
\beq
C_{k(x)}(\tau)=\frac{1}{a^{k(x)}} f(\tau)=x^\gamma f(\tau),
\eeq
so that Eq.~(\ref{scalingckcont}) takes the form
\beq
D(x,\tau)\approx \frac{1}{x^{1-\gamma}} f(\tau).
\eeq
This is indeed the expected scaling solution (cf. Eq.~(\ref{scal-spect})). 
The continuous distribution $D(x,\tau)$ is plotted in Fig.~\ref{scalingA}, where the scaling behavior is clearly seen. As we have shown in the previous section, this scaling regime sets in for $x\lesssim x_s(\tau)\sim\tau^2$, with $x_s$ separating  the single and multiple branching regimes (for $\tau\ll 1$).
To see how this scale emerges in the present model, let us recall that the coefficients $C_k$ take the form of a series in $e^{-a\tau}$, with $a=\sqrt{2} >1$, that is
\beq
C_k=\sum^k_{m=0} B(m,k) \, \rme^{-a^m \tau}. 
\eeq
where the coefficients $B(m,k)$ can be easily inferred from Eq.~(\ref{CkBwos}). In the continuum limit the sum over the modes $m$ turns into an integral over $x'=1/2^m$, 
\beq
C_{k(x)}\approx \frac{1}{\ln 2}\int^1_x \frac{\rmd x'}{x'} B[m(x'),k(x)] \rme^{-\frac{\tau }{\sqrt{x'}}}. 
\eeq
For a given $\tau\ll 1$, the dilue regime corresponds to the region where $\sqrt{x}\lesssim \tau$: there one can expand the exponential. Such an expansion is no longer possible when when $x <x_s(\tau)$: then the argument of the exponential is of order unity for all values of $x^\prime$ such that $\sqrt{x^\prime}<\tau$.

%%%%%%%%%%%%%%%%%%%%%%%%%%%%%
\subsection{Constant kernel}
%%%%%%%%%%%%%%%%%%%%%%%%%%%%%%
The previous models capture the main qualitative features of the cascades that we have introduced in the previous sections. In order to see the influence of the kernel ${\cal K}(z)$ on these results, we now examine the case of a flat kernel
\beq
{\cal K}(z)=1,
\eeq
for which analytic solutions can also be obtained. We shall still consider the two cases of model A in which $t_\ast(x)=t_\ast$ is constant, and model B where $t_\ast(x)=t_\ast\sqrt{x}$. 
As before, we set $\tau=t/t_\ast$.  As we shall see, the qualitative features of the cascades of both models are not much altered by the fact that the flat kernel allows at each splitting energy to be transported all the way down to zero. This is in line with the remark made in Sect.~\ref{sect:general}, which suggests that only a sufficiently singular kernel would lead to a substantial modification of the cascade properties.

\subsubsection{Model A}
\begin{figure}[h]

		\includegraphics[width=6.8cm]{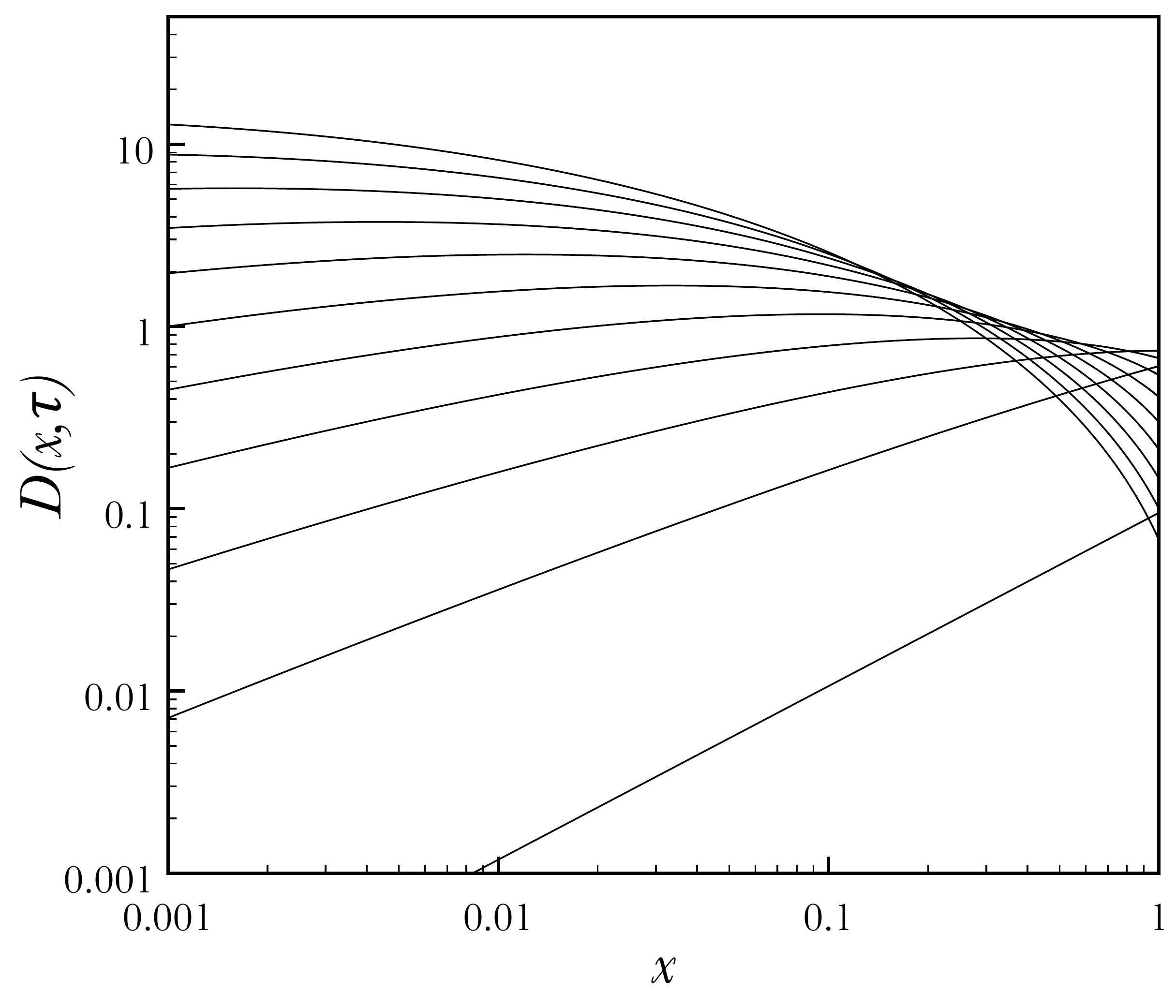} \includegraphics[width=6.8cm]{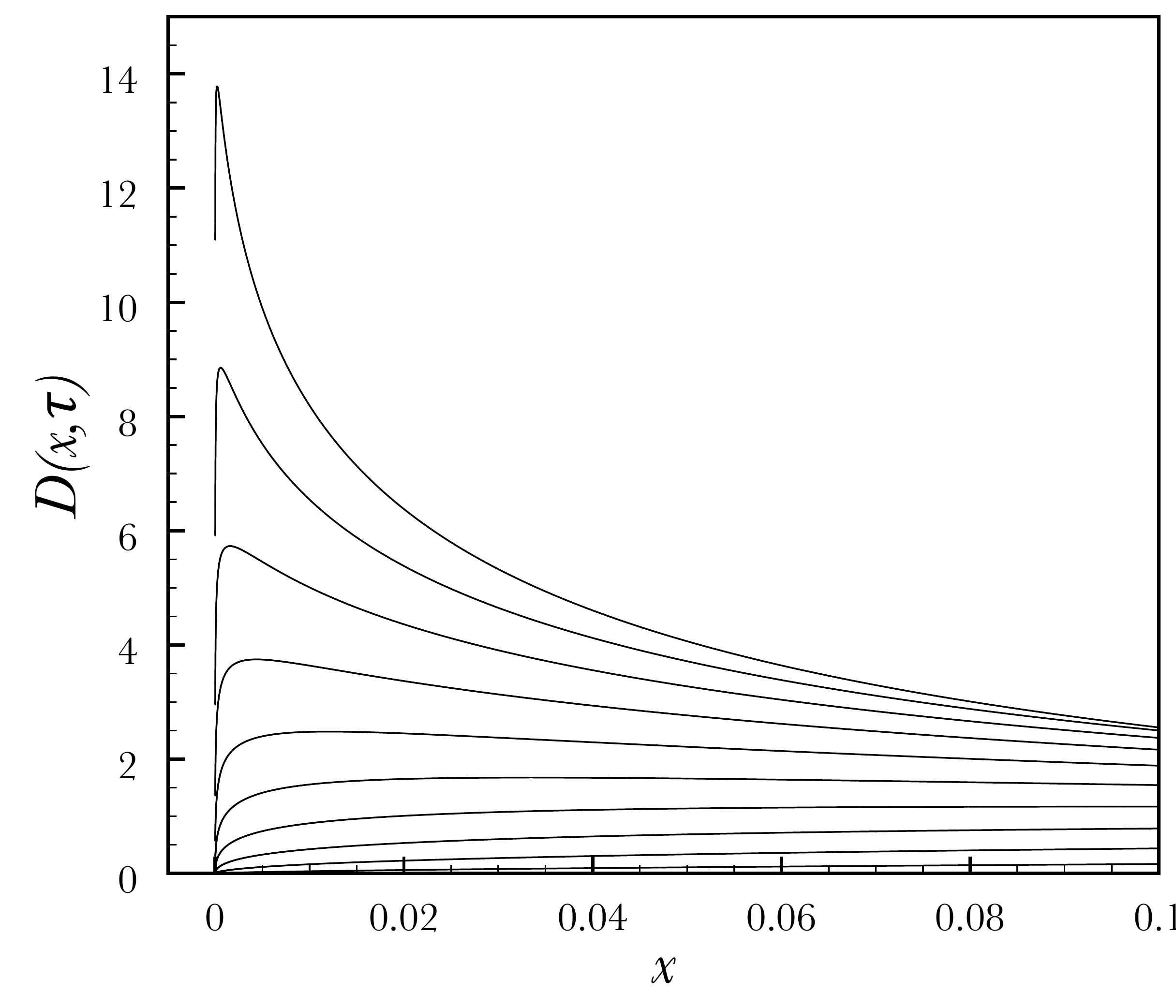} 
		\caption{Model A with a flat kernel. Left panel: the function $D(x,\tau)$ in Eq.~(\ref{DmodelAflat}), with $\tau$ increasing from bottom to top, $\tau=0.1,1,2,3,4,5,6,7,8,9,10$. Right panel: the same function $D(x,\tau)$  in a linear plot, exhibiting the dynamically generated infrared cutoff, in agreement with Eq.~(\ref{dyncutoff}). }
		\label{fig:flat_A_B}

\end{figure}

The solution is obtained in \ref{flat-A} and reads
\beq\label{DmodelAflat}
D(x,\tau)=x\sqrt{ \frac{\tau}{\ln\frac{1}{x}}}\, I_1\left( 2\sqrt{ \tau\ln\frac{1}{x}}\right)\,\rme^{-\frac{1}{2}\tau}, 
\eeq
where $I_1(x)$ is a modified Bessel function of the first kind. The exponential decay is slower than in the strict democratic branching case: this reflects the fact that very asymmetric branchings are now possible, and these leave a large fraction of the energy in the vicinity of the leading particle, which as a result is seen to decay at a smaller rate. At large $x$ and not too long time, $D(x,\tau)\sim x\tau$, a linear behavior that is clearly visible in Fig.~\ref{fig:flat_A_B}, left panel. One also observes on this figure the slow build up of the scaling solution (visible around $x\sim 0.1$). Another important feature of the solution is the  generation of a dynamical infrared cutoff. This is visible on Fig.~\ref{fig:flat_A_B} (right panel), and can be inferred from the asymptotic behavior of the distribution when $x\ll \rme^{-\tau}$, Eq.~(\ref{asympAflat}),  quite analogous to that of the DGLAP cascade (cf. Eq.~(\ref{DGLAPas0})), 
\beq\label{dyncutoff}
D(x,\tau) \approx \left(\frac{1}{x}\right)^{\sqrt{\frac{2\tau}{\ln\frac{1}{x}}}-1}.
\eeq
This behavior of $D(x,\tau)$ as $x\to 0$ is what keeps  the integral of the spectrum finite. As time goes on,  the energy injected in the system remains in the spectrum, the flow vanishing at $x=0$ (see \ref{flat-A}).

The same property holds when a source   ${\cal A}\delta(1-x)$ is added. In this case,  the solution converges slowly towards a stationary solution 
\beq
D_\text{st}(x)=\frac{4{\cal A}}{x}, 
\eeq
where we used the fact that $v=1/4$ for a flat kernel (see Eq.~(\ref{v-flow})). As already emphasized, this solution is reached only for values of $x$  above the dynamical infrared cutoff $x_{\rm min}$ (see Eq.~(\ref{xmin}) and the discussion at the end of Sect.~\ref{sect:general}).

%%%%%%%%%%%%%%%%%%%%%%%%%%%%%%%%%%
\subsection{Model B }\label{DGLAPsolution-2}
%%%%%%%%%%%%%%%%%%%%%%%%%%%%%%%%%%%

\begin{figure}[h]
\begin{center}
		 \includegraphics[width=6.5cm]{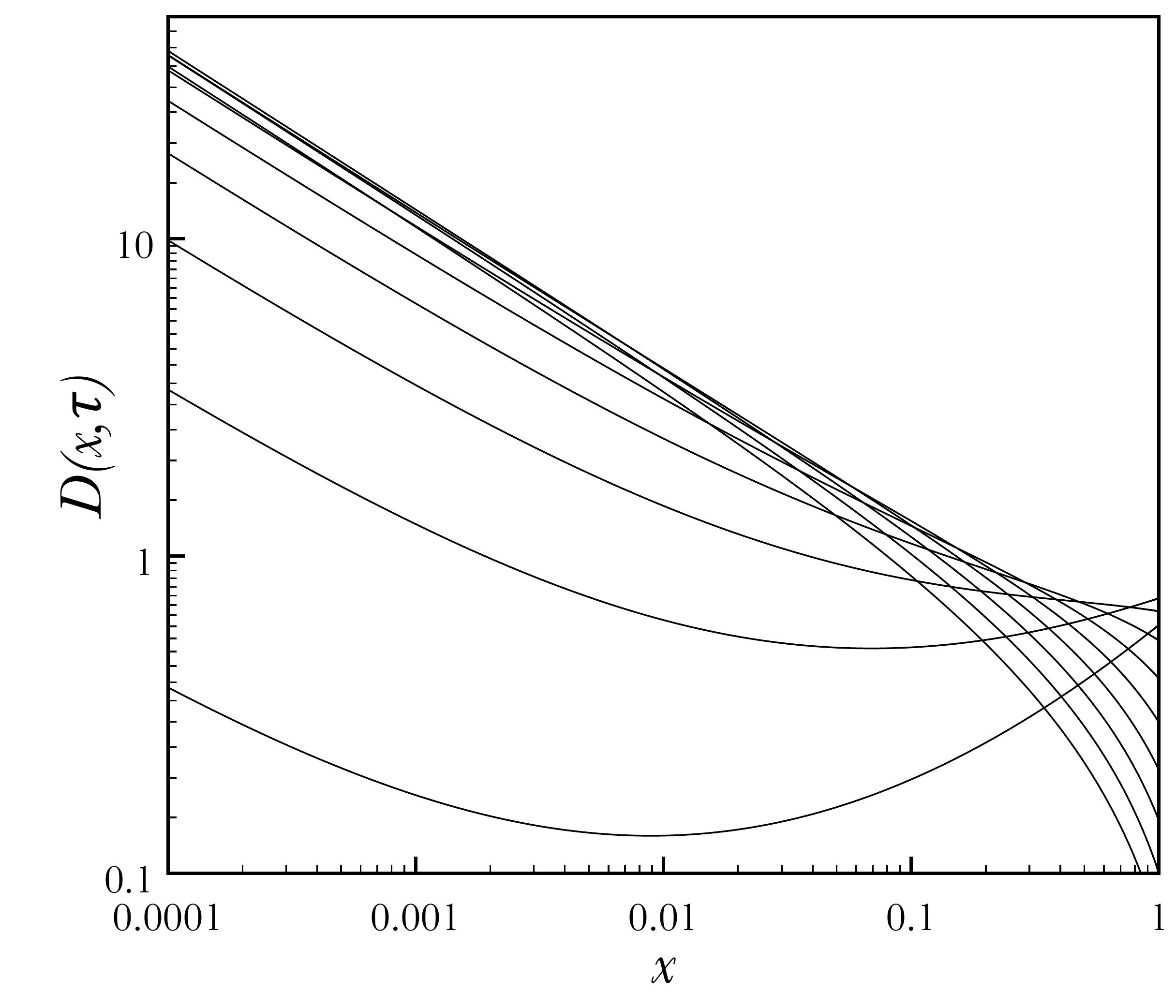}  \includegraphics[width=6.5cm]{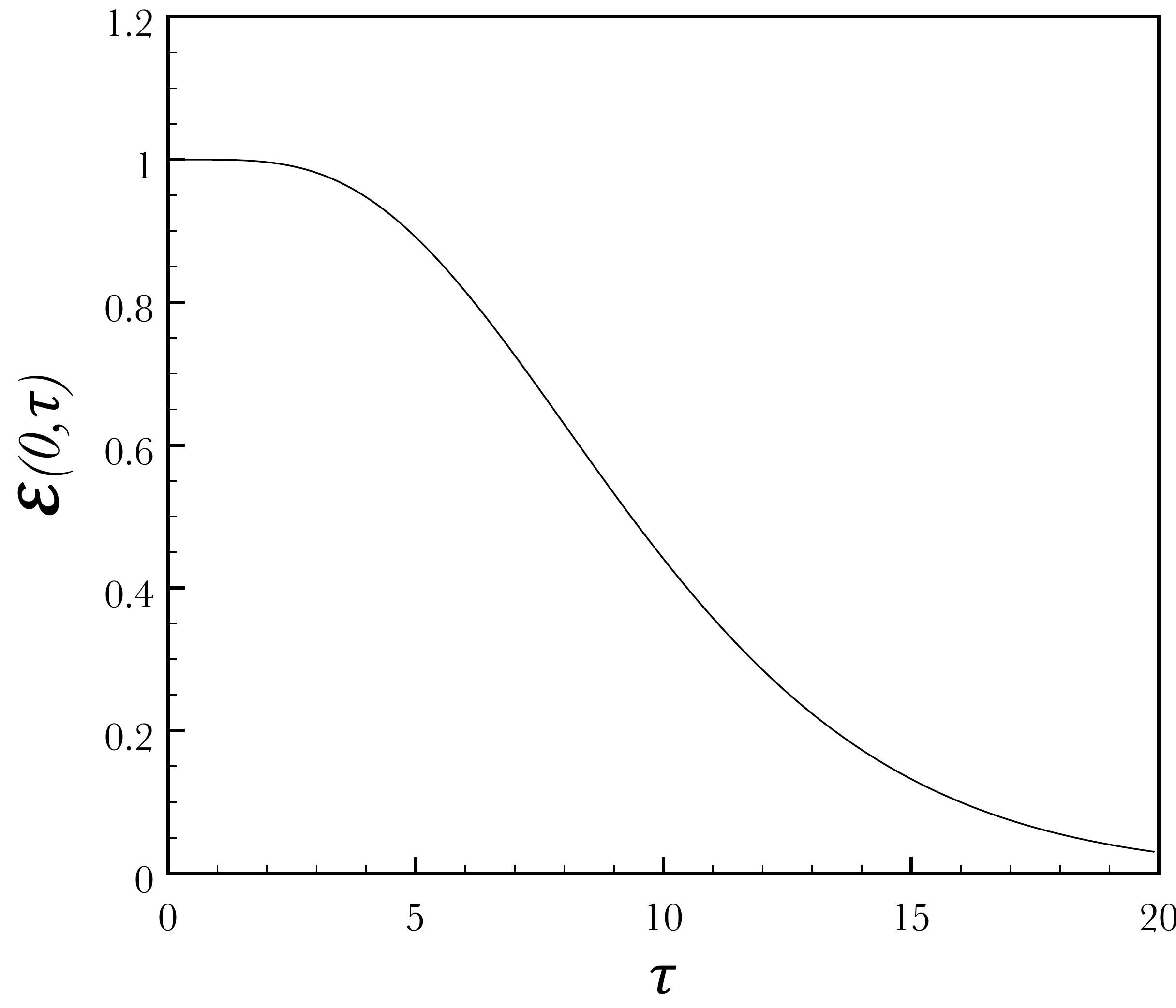}
		\caption{Model B with a flat Kernel. The left panel shows the evolution with time of the solution (\ref{solution-B-flat1}), with $\tau=0.1,0.5,1,2,3,4,5,6,7,8,9,10$ from bottom to top. The right panel shows how the total energy (\ref{energytotale}) contained in the spectrum decreases in time. }
		\label{energy}
\end{center}
\end{figure}

The solution corresponding  to model B is obtained in \ref{flat-B}. It can be written in the form
\beq\label{solution-B-flat1}
D(x,\tau)= g(x,\tau)+\,\rme^{-\frac{1}{2}\tau}  \delta(1-x).
\eeq 
where 
\beq\label{solution-B-flat}
g(x,\tau)=f_{-1/2}(\tau)\, \frac{1}{ \sqrt{x} }+ f_0(\tau) +f_{1/2}(\tau) \,\sqrt{x}+f_{1}(\tau) \,x,
\eeq
and the various $f_i(\tau)$ are given in \ref{flat-B}.
At early times, this solution behaves very much like that of model A, 
\beq
D(x,\tau)\simeq x \tau.
\eeq
However, as can be seen on Fig.~\ref{energy}, right panel, a scaling behavior in $1/\sqrt{x}$ quickly develops at small $x$. The explicit calculation of \ref{flat-B} indicates that the scaling solution is of the form 
\beq
D(x,\tau)\simeq \frac{\tau^4}{192\sqrt{x}} \,\rme^{-\frac{1}{2}\tau}.
\eeq
The typical value of $x$ at which the transition occurs is therefore $x_s\sim \tau^2$ (see the discussion at the end of Sect.~\ref{sect:general}). 

The integral of the spectrum reads:
\beq\label{energytotale}
\int_0^1\rmd x \,D(x,\tau)=2 f_{-1/2}(\tau)+f_0(\tau)+\frac{2}{3} f_{1/2}(\tau)+\frac{1}{2} f_1(\tau).
\eeq
In contrast to model A, this is a decreasing function of time, as can be seen in Fig.~\ref{energy}, right panel. Since the energy is conserved at each step of the cascade, this implies that the energy flows all the way down to $x=0$ where a condensate develops.  The solution in the presence of a source is determined in \ref{flat-B}. 
The stationary solution  in this case reads
\beq
D_\text{st}=\frac{4{\cal A}}{\sqrt{x}}, 
\eeq
and apart from a different $x$ dependence, is similar to that of model A.

%%%%%%%%%%%%%%%%
\section{Conclusions}
%%%%%%%%%%%%%%%%%
The present analysis clarifies the distinct properties that characterize the medium-induced, or BDMPS cascade from the vacuum DGLAP cascade. In the first case, the leading parton is essentially on-shell, and radiation is induced by the multiple collisions that the leading parton and its offsprings undergo with the medium constituents. In the second case, the leading parton is off-shell, and the evolution parameter is the virtuality of the emitted radiation. We have seen that the two types of cascades may be viewed, formally, in a broader context where the key quantity that differentiates them appears to be a suitably defined branching time, i.e., the typical time it takes a gluon to branch quasi-democratically into two gluons with comparable fractions of the initial energy. In the DGLAP cascade, this time is constant, while it decreases along the BDMPS cascade. As a result, it takes a finite  time to transport a finite amount of energy from the leading particle to very soft gluons, while the corresponding time is infinite in the DGLAP cascade. This property is at the origin of the scaling behavior of the energy distribution in the BDMPS cascade, which exhibits indeed many properties that are reminiscent of wave turbulence. The DGLAP cascade can in fact be viewed as a limiting case of the BDMPS cascade, but since the relevant transport time scale is infinite, it takes an infinite amount of time to reach the scaling solution, which is therefore not observed.\\

In this paper, we have focused on the energy flow, but the present study has a broader scope. In particular,  as shown recently, the same characteristic features  of the BDMPS cascade as exhibited here play an essential role in determining the angular distribution of the produced radiation \cite{Blaizot:2014ula,Kurkela:2014tla,Blaizot:2014rla}.  We have discussed only the ideal case where the BDMPS mechanism is allowed to work all the way down to zero energy. This is of course an idealization. In an actual cascade propagating in a quark-gluon plasma, the energy would not accumulate at zero energy, but rather at the typical scale where dissipation would occur, namely the temperature of the system, that  also plays the role of the Bethe-Heitler momentum cutoff. While the generic features of the BDMPS cascade that we have discussed here are, to a large extent, independent of this lower cutoff, the question of how precisely the energy of such a cascade dissipates its energy in the medium is an interesting one that deserves further studies (for a recent discussion see \cite{Kurkela:2014tea}). \\

\noindent{\bf Acknowledgements}

We acknowledge fruitful discussions with E. Iancu at an early stage of this work. 
This research is supported by the European Research Council under the Advanced Investigator Grant ERC-AD-267258.

%%%%%%%%%%%%%%%%%%%
%%%%%%%%%%%%%%%%%%%%%
%%%%%%%%%%%%%%%%%%%%%
\appendix

\section{Exact solutions: BDMPS-like cascade}
\label{exactlargexc}

In order to solve Eq.~(\ref{Dfin2}), it is convenient to perform a change of variable that gives the gain and loss terms a more symmetrical form. 
In the gain term, we set $\xi=x/z$ and in the loss term, $\xi=xz$. 
A simple calculation shows that, in terms of these new  variables, the equation reads
\beq\label{main}
\del_\tau D(x,\tau)=\int_x^1\rmd \xi\, P(x,\xi)D(\xi,\tau)-D(x,\tau)\int_0^x \rmd\xi P(\xi,x),
\eeq
with 
\beq
P(x,\xi)\equiv \frac{x}{\xi^{3/2}} {\cal K}(x/\xi)=\sqrt{\frac{\xi}{x}}\frac{1}{(\xi-x)^{3/2}}\,.
\eeq
Note that the singularities at $z=1$ in the original formulation have been moved to $\xi=x$ in both the gain and the loss term. These singularities cancel each other when both terms are added. 

In the following we shall consider first the case of a large medium, $L> t_\ast$, i.e., $\omega_c< E$, and next the case $\omega_c\ll E$.

%%%%%%%%%%%%%%%%%%%%%%%%
 \subsection{The exact solution for  $x_c>1$}
%%%%%%%%%%%%%%%%%%%%%%%%
We consider first the case of a large medium, $L> t_\ast$, i.e., $\omega_c>E$ or $x_c=\omega_c/E>1$. 
In order to calculate separately the gain and the loss terms, we may introduce a regularized splitting function $P_\epsilon(x,\xi)$
\beq
P_\epsilon(x,\xi)=\sqrt{\frac{\xi}{x}}\,\frac{1}{(\xi-x+\epsilon)^{3/2}}.
\eeq
Let us  analyze the loss term with this regularized splitting function. The integral over $\xi$  can be calculated explicitly
 \beq
\int^x_0 d\xi \sqrt{\frac{x}{\xi}}\frac{1}{(x-\xi+\epsilon)^{3/2}}=\frac{1}{\sqrt{\epsilon}}\frac{2x}{x+\epsilon}\approx \frac{2}{\sqrt{\epsilon}}-\frac{2\sqrt{\epsilon}}{x}+{\cal O}(\epsilon^{3/2}).
\eeq
Note that apart from the fist term that is divergent when $\epsilon\to 0$ all subleading terms vanish (for any finite $x$). Hence, in these variables,  the sole effect of the loss term is to remove the singularity of the gain term. Accordingly we shall write the loss term symbolically  as
\beq\label{lossa}
\int^x_0 d\xi P(\xi,x)\longrightarrow\int_0^\infty \frac{dz}{z^{3/2}}.
\eeq 

In order to proceed we now  introduce the following rescaled distribution 
$F(y,\tau)\equiv\sqrt{x}\,D(x,\tau)$, with $y=1-x$.
This has the advantage of removing  the expected singular behavior of $D(x)$ at small $x$.
We get then
\beq\label{evol7a}
\frac{\del}{\del \tau} F(y,\tau)= \int_0^{y} \rmd\xi\frac{1}{(y-\xi)^{3/2}} \,F\left(\xi,\tau\right)- F(y,\tau)\,\int_y^1 \frac{\rmd\xi}{\sqrt{1-\xi}}\frac{1}{(\xi-y)^{3/2}}.\nn
\eeq
At this point, it is convenient to extend the domain where $F(y,\tau)$ is defined, from $[0,1]$ to $[0,\infty]$, which amounts to extend the definition of $D(x)$ to unphysical (negative) values. The motivation for doing so is that it allows us to  take a Laplace transform. Defining the Laplace transform as 
\beq
\tilde F(\alpha,\tau)=\int_0^\infty dy\, e^{-\alpha y}\,F(y,\tau),
\eeq
we get, for the gain term,
\beq
\int_0^\infty dy\, \rme^{-\alpha y}\, \int_0^{y} \frac{d\xi}{(y-\xi)^{3/2}} \,F\left(\xi,\tau\right)=\int_0^\infty \rmd\xi\, \rme^{-\alpha \xi}F\left(\xi,\tau\right)\int_0^\infty\frac{\rme^{-\alpha z}}{z^{3/2}}.\nn
\eeq
In this manipulation, the singularity at $\xi=y$ has been moved to $z=0$, where it takes exactly the same form as that of the loss term, Eq.~(\ref{lossa}), with which it  therefore cancels when gain and loss terms are added. We  then arrive at the finite equation for the Laplace transform $\tilde F(\alpha, \tau)$
\beq\label{evol8}
\frac{\del}{\del \tau} \tilde F(\alpha,\tau)= \int_0^\infty \frac{dz}{z^{3/2}} \left(e^{-\alpha z}-1\right) \,\tilde F(\alpha,\tau)=-2\sqrt{\pi \alpha} \,\tilde F(\alpha,\tau).
\eeq
For a single gluon, the initial condition  is $F(y,\tau=0)=\delta(y)$, which yields $\tilde F(\alpha,\tau=0)=1$.
 After integrating (\ref{evol8}) over $t$ and inverting the Laplace transfom, the solution of (\ref{evol7a}) reads
\beq\label{Fsol1}
F(y,\tau)=\int_{c-i\infty}^{c+i\infty} \frac{d\alpha}{2\pi i}\, e^{\alpha y-2\sqrt{\pi\alpha} \tau}=F(y,\tau)=\frac{\tau}{y^{3/2}}\, \exp\left(-\pi\frac{\tau^2}{y}\right).
\eeq
It follows that, in the physical domain,  $x=1-y$, $0<x<1$,
\beq\label{Gsol2Ap}
D(x,\tau)=\frac{\tau}{\sqrt{x}\,(1-x)^{3/2}}\, \exp\left(-\pi\frac{\tau^2}{1-x}\right)\,.
\eeq

%%%%%%%%%%%%%%%%%%%%%%
\subsection{Solution for $x_c\ll 1$}
\label{exactsmallxc}
%%%%%%%%%%%%%%%%%%%%%%

When $x_c\ll 1$, the emission processes are sensitive to the upper cutoff on the BDMPSZ spectrum (recall that this spectrum falls very rapidly with the energy $\omega$ of the radiated gluon when $\omega\gtrsim \omega_c=\hat q t^2$). We implement this suppression with a sharp cutoff at $\omega= \omega_c$, and modify the splitting function accordingly as follows 
\beq
P(x,\xi)\longrightarrow P(x,\xi)\,\theta_c(x,\xi),\qquad \theta_c(x,\xi)\equiv\theta(x_c-x)+\theta(x_c-\xi+x),\nn
\eeq
and the same for $P(\xi,x)$. It is also convenient to divide the distribution $D(x,\tau)$ into a soft component, $D_{\rm s}(x,\tau)=\theta(x_c-x)D(x,\tau)$, and a hard component, $D_{\rm h}(x,\tau)=\theta(x-x_c)D(x,\tau)$. As we shall see shortly, when $x_c\ll 1$, the  hard component obeys a closed, homogeneous, equation with initial condition
$D_h(x,\tau=0)=\delta(1-x)$, while the soft component obeys an inhomogeneous
equation where the hard component plays the role of a source. Note that the two components are separated by a gap, $x\in [x_c,1-x_c]$, where $D(x,\tau)=0$.

Let us consider first the hard component, that is the region $1-x_c\le x\le 1$. In this region, there is no constraint on the gain term, since  $\xi-x<x_c$. There is a constraint on the loss term however. We write this term as follows:
$
\int_0^x \rmd \xi P(\xi,x)-\int_{x_c}^{x-x_c}P(\xi,x).
$
An elementary calculation yields
\beq
\int_{x_c}^{x-x_c}P(\xi,x)=\frac{2}{\sqrt{x}}\frac{x-2x_c}{\sqrt{x}\sqrt{x-x_c}}=\frac{2}{\sqrt{x}}+{\cal O}\left(\frac{1}{\sqrt{x_c}}\frac{x_c}{x}\right),
\eeq
where in the last step, we have used $x_c\ll 1 $. In this limit, the correction to the loss term is independent of $x$ and the solution  $D_{\rm h}(x,t)$ reads simply
\beq
D_{\rm h}(x,\tau)\approx \frac{\tau}{(1-x)^{3/2}}\,{\rm e}^{-\frac{\pi \tau^2}{1-x}+\frac{2\tau}{\sqrt{x_c}} }.
\eeq

Turning now to the soft region ($x\le x_c$), we note that the cutoff affects there only the gain term. We write
\beq
\del_\tau D_{\rm s}(x,\tau)&=&\int_{1-x_c}^1 \rmd \xi P(x,\xi) D_{\rm h}(\xi,\tau)+\int_x^{x_c} \rmd \xi P(x,\xi) D_{\rm s}(\xi,\tau)\nn &-&\int_0^x\rmd\xi P(\xi,x)D_{\rm s}(x,\tau).
\eeq
The first term on the right hand side plays the role of a source term. It can  be calculated explicitly:
\beq
S(x,\tau)=\frac{1}{\sqrt{x}} \,{\rm erfc}\left(\sqrt{\frac{\pi}{x_c}\tau}\right)\,{\rm e}^{2\tau/\sqrt{x_c}}.
\eeq
By rescaling the variables, setting $\xi=ux_c$, $x=z x_c$, $\tau=\sqrt{x_c}\tau_c$, one can put the equation for $D_{\rm s}$ in the form
\beq
\del_{\tau_c} D_{\rm s}(z,\tau_c)=\frac{1}{\sqrt{z}} {\rm erfc}\left(\sqrt{\pi}\tau_c\right)\,{\rm e}^{2\tau_c}&+&\int_z^1\rmd uP(z,u)D_{\rm s}(u,\tau_c)\nn &-&\int_0^z \rmd u P(u,z)D_{\rm s}(z,\tau_c).
\eeq
This is the original equation (see Eq.~(\ref{main}))  with a source term. Again, this can be solved via Laplace transform, using manipulations similar to those described earlier. We only write here the final solution in terms of the original variables,
\beq
D_{\rm s}(x,\tau)=\frac{1}{\sqrt{x}}\int_0^{\tau} \rmd \tau' {\rm erfc}\left(\sqrt{\frac{\pi}{x_c}}\tau'\right)\,{\rm e}^{2\tau'/\sqrt{x_c}}\,{\rm erfc}\left(\sqrt{\frac{\pi}{x_c-x}}(\tau-\tau')\right).\nn
\eeq
Note that in the limit $x\ll x_c$ and for $\tau=\tau_{\rm max}=\bar\alpha \sqrt{x_c}$, one can expand around $\tau=0$. We get
\beq
D_{\rm s}(x,\tau)\approx \bar\alpha \sqrt{\frac{x_c}{x}}\left[ 1-\bar\alpha-\frac{2}{3}\bar\alpha^2 +{\cal O}(\bar\alpha^3)  \right].
\eeq
One recovers corrections that can be obtained by a perturbative calculation \cite{Fister:2014zxa}.
More generally, the structure of the cascade is that of a peak localized near $x=1$, feeding soft modes at small $x\lesssim x_c$. The main features of the resulting cascade of soft modes are quite similar to those of the general cascade that we have detailed earlier in this section. \\

 We note finally, that although we have focused on the cascade generated by a gluon, a similar analysis can be made for the case where the leading particle is a quark. The corresponding energy distribution is obtained by using the following simplified form for the kernel 
\beq\label{gker2b}
{\cal K}(z)\approx\frac{1}{z^{1/2}(1-z)^{3/2}}\,,
\eeq
It is easy to check that the quark distribution is related to the gluon distribution as follows  $D_q(x,t)=xD_g(x,t)$, since the loss term stays unchanged. Hence, 
\beq\label{Qsol1}
D_q(x,t)=\frac{\sqrt{x}\,t}{(1-x)^{3/2}}\, \exp\left(-\pi\frac{t^2}{1-x}\right)\,.
\eeq

%%%%%%%%%%%%%%%%%%%%%%%%%%%%%%%%%%
\section{Solution of the DGLAP cascade}\label{DGLAPsolution}
%%%%%%%%%%%%%%%%%%%%%%%%%%%%%%%%%%%

In this Appendix, we write the solution of the  DGLAP evolution equation, Eq.~(\ref{DGLAP}) in  Mellin space. We define  the Mellin transform of the energy distribution, and its inverse, respectively as 
\beq
\tilde D(\nu,t)=\int_0^1 \rmd x x^{\nu-1}D(x,t), \qquad
D(x,t)=\int_{c-i\infty}^{c+i\infty} \frac{\rmd\nu}{2\pi i} x^{-\nu}\tilde D(\nu,t),
\eeq
where the integration contour in the inverse transform runs parallel to the imaginary axis, to the right of the poles of the integrand. 
By taking the Mellin transform of Eq.~(\ref{DGLAP}), and recognizing the integral representation of the Digamma function,
\beq
\int _0^1 \frac{1-z^{\nu}}{1-z}\rmd z\,-\gamma=\psi(\nu) \equiv \frac{\Gamma'(\nu)}{\Gamma(\nu)},
\eeq  
one obtains 
\beq\label{Mellin-DGLAP}
\frac{\del }{\del t} \tilde D(\nu,t) =-\left( \psi(\nu)+\gamma \right) \tilde D(\nu,t),
\eeq 
where $\gamma$ is the Euler constant. It is straightforward to solve Eq.~(\ref{Mellin-DGLAP}), with the initial condition $\tilde D(\nu,0)=1$,  the Mellin transform of $D(x,0)=\delta(1-x)$. The solution reads
\beq\label{DGLAP-sol-Mellin}
\tilde D(\nu,t) =\exp\left[-(\psi(\nu)+\gamma)t \right].
\eeq
Transforming back to $x$ space, we write the solution of the DGLAP equation as 
\beq\label{DGLAP-solb}
D(x,t)=\int_{c-i\infty}^{c+i\infty} \frac{\rmd\nu}{2\pi i}  \exp\left[-(\psi(\nu)+\gamma)t+\nu\ln\frac{1}{x}\right].
\eeq
In general it is difficult to perform the integration over $\nu$ in Eq.~(\ref{DGLAP-sol}). But in some limits one can perform analytic evaluations which are be sufficient for the purpose of this work. This is the case in particular for the small $x$ limit where one can use the saddle-point approximation.  The argument of the exponential has a minimum when
\beq
\psi'(\nu_s) =\frac{\ln 1/x}{t}.
\eeq  
When $\ln\frac{1}{x} \gg t $, $\psi(\nu) \approx -\nu^{-1}$ and $\psi'(\nu) \approx \nu^{-2}$, so that $
\nu_s= \sqrt{\frac{t}{\ln 1/x}}$.
Hence we get
\beq\label{DGLAP-sol-approx1}
D(x,t)&\simeq & \exp\left[-\gamma t+2 \sqrt{t\,\ln\frac{1}{x} }\right] \int_{c-i\infty}^{c+i\infty} \frac{\rmd\nu}{2\pi i}  \exp\left[ \frac{1}{2}\psi''(\nu_s)t (\nu-\nu_s)^2\right],\nn
&=& \frac{1}{2}\left(\frac{ t}{\pi^2 \ln^3 1/x }\right)^{1/4}\exp\left[-\gamma t+2 \sqrt{t\,\ln\frac{1}{x} }\right].
\eeq
Focusing on the leading behavior,  we get
\beq\label{DGLAPas0}
D(x,t)\approx \left(\frac{1}{x}\right)^{2\sqrt{\frac{t}{\ln 1/x}}}.
\eeq
This is the standard Double Logarithmic limit of DGLAP. Note that the $t$ and $x$ dependence of the energy distribution do not factorize. Obviously, since $\ln 1/x \gg t$, the spectrum is integrable at small $x$. This can be verified by an explicit calculation. 
From Eq.~(\ref{DGLAP-sol-Mellin}) we have indeed 
\beq
E=\int_0^1 \rmd x D(x,t) = \tilde D(1,t) =1.
\eeq

%%%%%%%%%%%%%%%%%%%%%
\section{Exact solutions: discrete models with democratic branching}\label{exactdemo}
%%%%%%%%%%%%%%%%%%%%%
In this Appendix, we provide details on the analytical solution of the the discrete models that are introduced in the main text (Section 4). Recall that in these models, the distribution $D(x,t)$ has support on a discrete set of points, $x=1/2^k$, the energy in each mode $k$ being denoted by $C_k(\tau)$. We set here $t_\ast=1$ so that $\tau=t$.
The coefficients $C_k$ satisfy the recursion relations (\ref{recusrionCk}), that we reproduce here for convenience
\beq\label{recusrionCkb}
\dot C_0={\cal A}-C_0,\qquad \dot C_k=C_{k-1}a^{k-1}-C_ka^{k} \quad (k\ge 1),\qquad  a\equiv 2^\gamma.\nn
\eeq
In the main text, we introduced two models, referred to respectively as model A and model B. In model A, $a=1$ ($\gamma=0$), which corresponds to a constant splitting rate along the cascade. In  model B, $a>1$ ($\gamma>0$), which implies an acceleration of the splittings along the cascade.

The initial condition depends on whether we consider the system with a source, ${\cal A}\ne 0$, or without a source, ${\cal A}=0$. In the first case, the initial condition is $C_{k}(t=0)=0$ for all $k$, while in the latter case, $C_{k\ge 1}(t=0)=0$, $C_0(t=0)=1$, with $C_0$ representing the energy of the leading particle. We also assume that at large time $C_k(t)$ vanishes or is bounded (we verify a posteriori that this is indeed the case).  We then solve the equations (\ref{recusrionCkb}) with a Laplace transform. We define, with a slight abuse of notation,
\beq
C_k(\omega)=\int_0^\infty \rmd t\, {\rme}^{-\omega t} C_k(t), \qquad C_k(t)=\int_{-i\infty}^{+i\infty} \frac{\rmd\omega}{2\pi}\, \rme ^{\omega t}\,C_k(\omega). 
\eeq
\\

Let us first analyze the case of a leading particle carrying initially all the energy, i.e., the case ${\cal A}=0$. By taking the Laplace transform of  Eq.~(\ref{recusrionCkb}) we obtain
\beq
(1+\omega) C_0(\omega)=1,\qquad \omega C_k(\omega)=a^{k-1}C_{k-1}(\omega) -a^k C_k(\omega) \quad (k\ge 1),\nn
\eeq 
where we have used $C_k(\infty)=0$ and $C_k(0)=\delta_{k0}$. The recursion relation can  be solved easily, with the result
\beq\label{CkLaplace}
C_k(\omega) =\prod_{m=1}^k\frac{a^{m-1}}{\omega+a^m}\,C_0(\omega),\qquad C_0(\omega)=\frac{1}{1+\omega}.
\eeq
The inverse Laplace transform then yields
\beq\label{inverseLaplace}
C_k(t)=\int_{-i\infty}^{i\infty} \frac{\rme ^{\omega t}}{1+\omega}\left( \prod_{m=1}^k\frac{a^{m-1}}{\omega+a^m} \right).
\eeq
\\

For model A, where $a=1$, the integrand in Eq.~(\ref{inverseLaplace}) has a multiple pole of order $k+1$ at $\omega=-1$. By closing the contour on the left half-plane ($t>0$) and picking up the contribution of the residue, one obtains easily
\beq\label{Poisson2}
C_k(t)=\frac{1}{k!}t^k \rme^{-t}.
\eeq
In the case of model B, where $a>1$, the integrand exhibits a series of distinct poles, $\omega=-1, -a, -a^2, \cdots, -a^k$. By closing the contour in the left half plane, and collecting the contributions of the residues one then gets, 
\beq\label{CkBwos}
C_0(t)&=&\rme^{-t},\nn C_{k\ge 1}(t)&=&\rme^{-t}\prod_{m=1}^k\frac{a^{m-1}}{a^m-1}+\sum_{m=1}^k\left\{\rme^{-a^m t} \frac{a^{m-1}}{1-a^m}   \left( \prod_{n\ne m}^k\frac{a^{n-1}}{a^n-a^m}  \right)\right\}.\nn
\eeq
Note that the exponential decay of $C_0(t)$, which represents the decay of the leading particle, is the same in both models. The way the population of the other modes ($k>0$)  varies with time depends  however on the model in a significant way. 

The Poisson distribution (\ref{Poisson2}) may be viewed as the limit of the general distribution (\ref{CkBwos}) when $a\to 1$.  In this limit, all poles at $-a,-a^2,\cdots, a^k$ merge together with the pole at $\omega=-1$, producing a pole of order $k+1$ at $\omega=-1$.  

\begin{figure}

		\includegraphics[width=6.5cm]{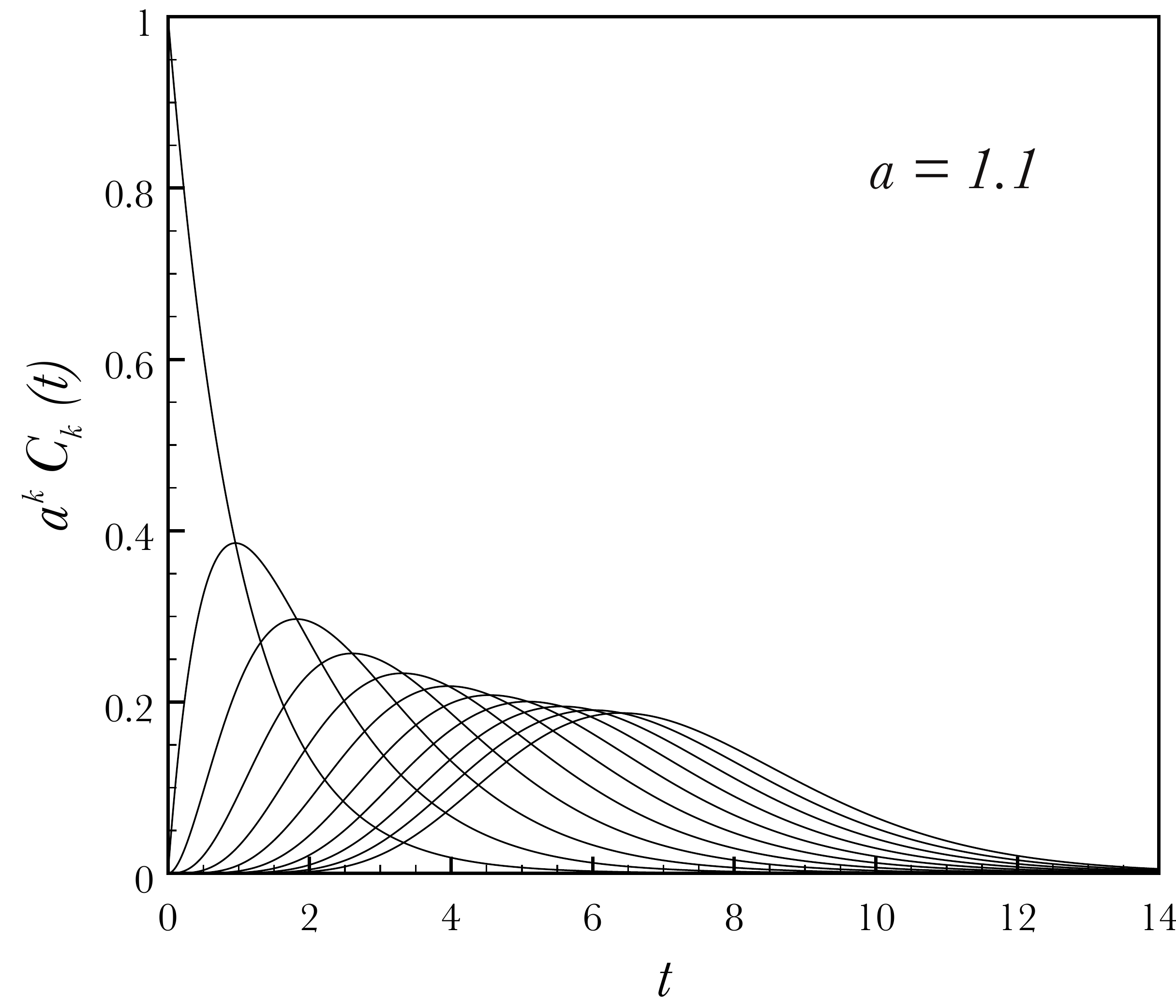}\includegraphics[width=6.5cm]{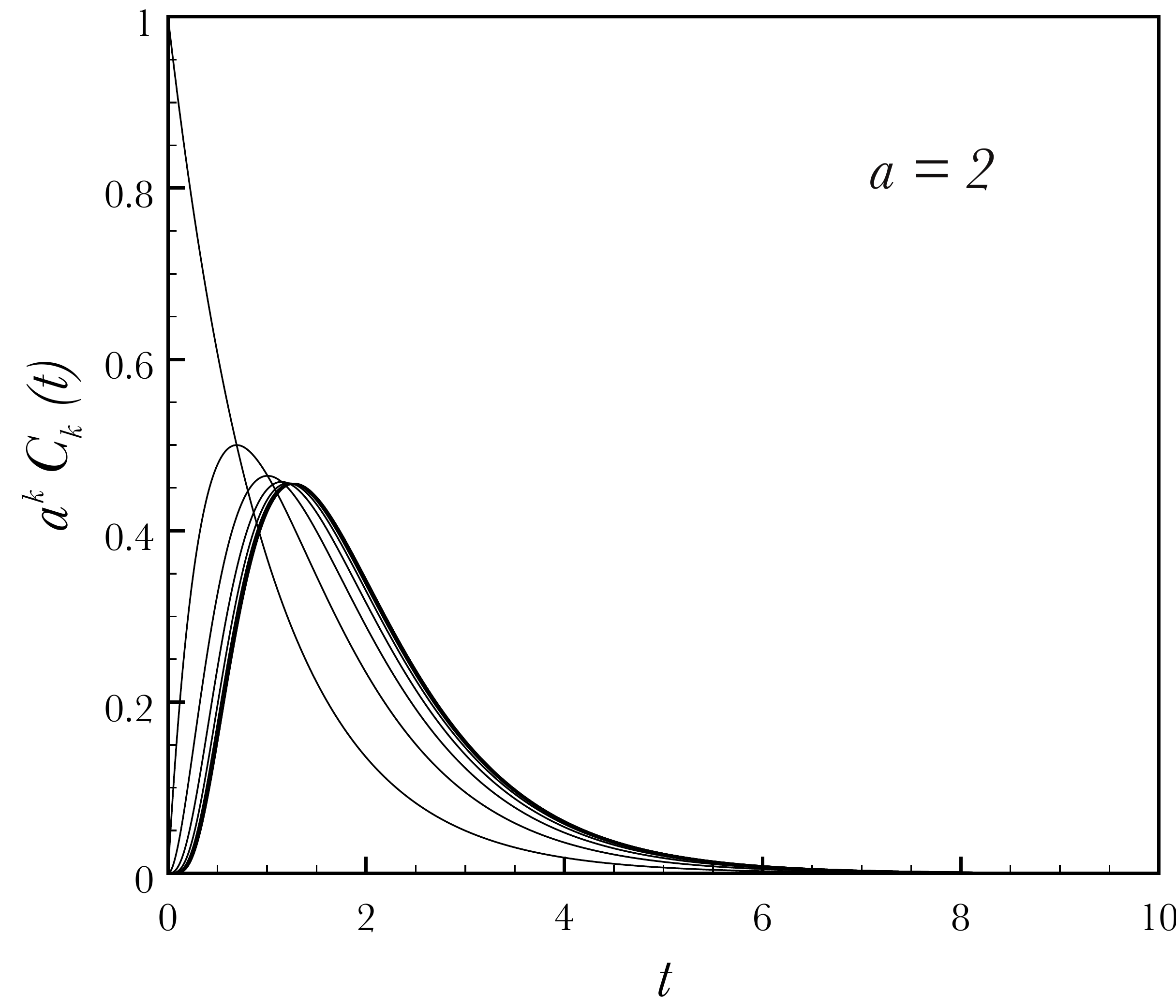} 
		\includegraphics[width=6.5cm]{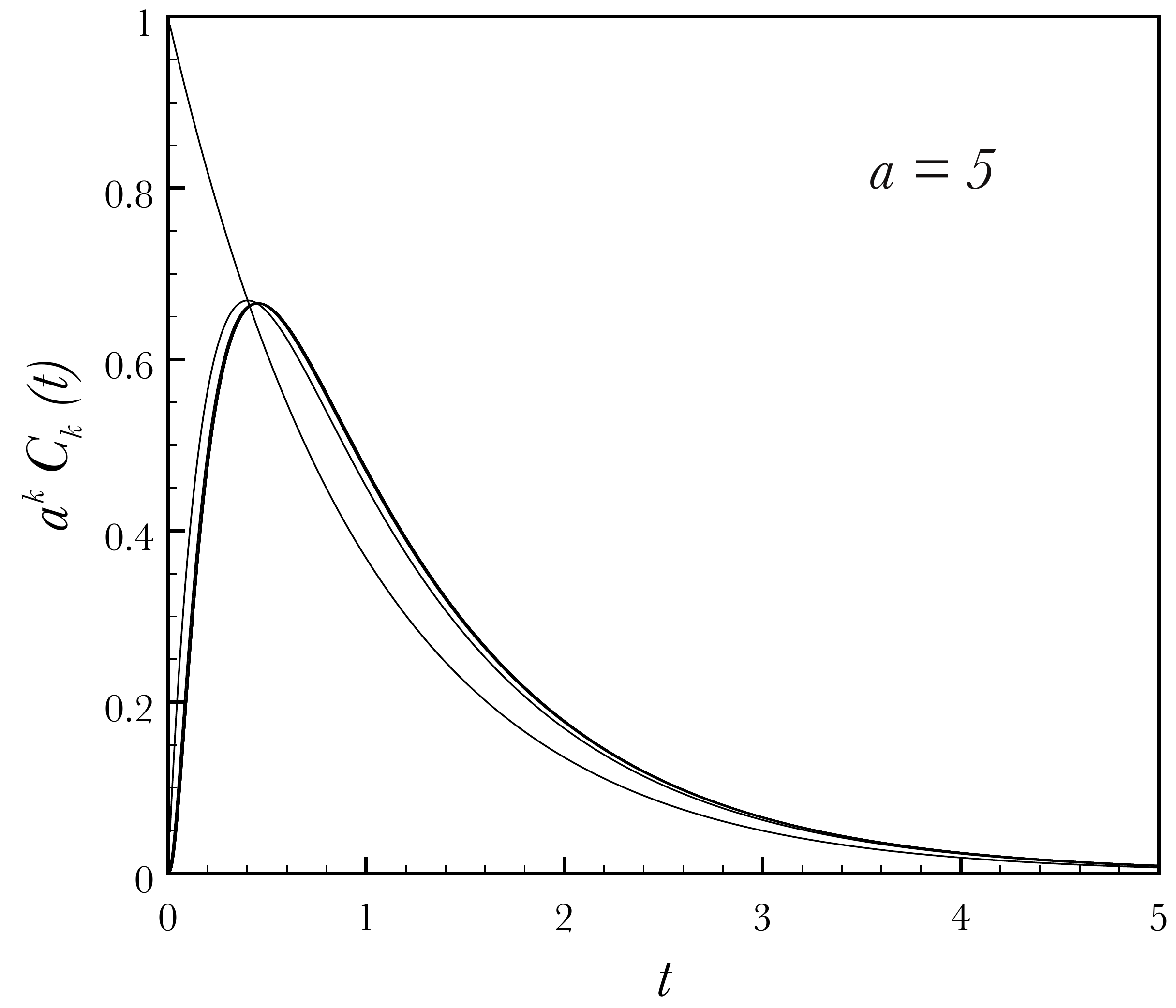}\includegraphics[width=6.5cm]{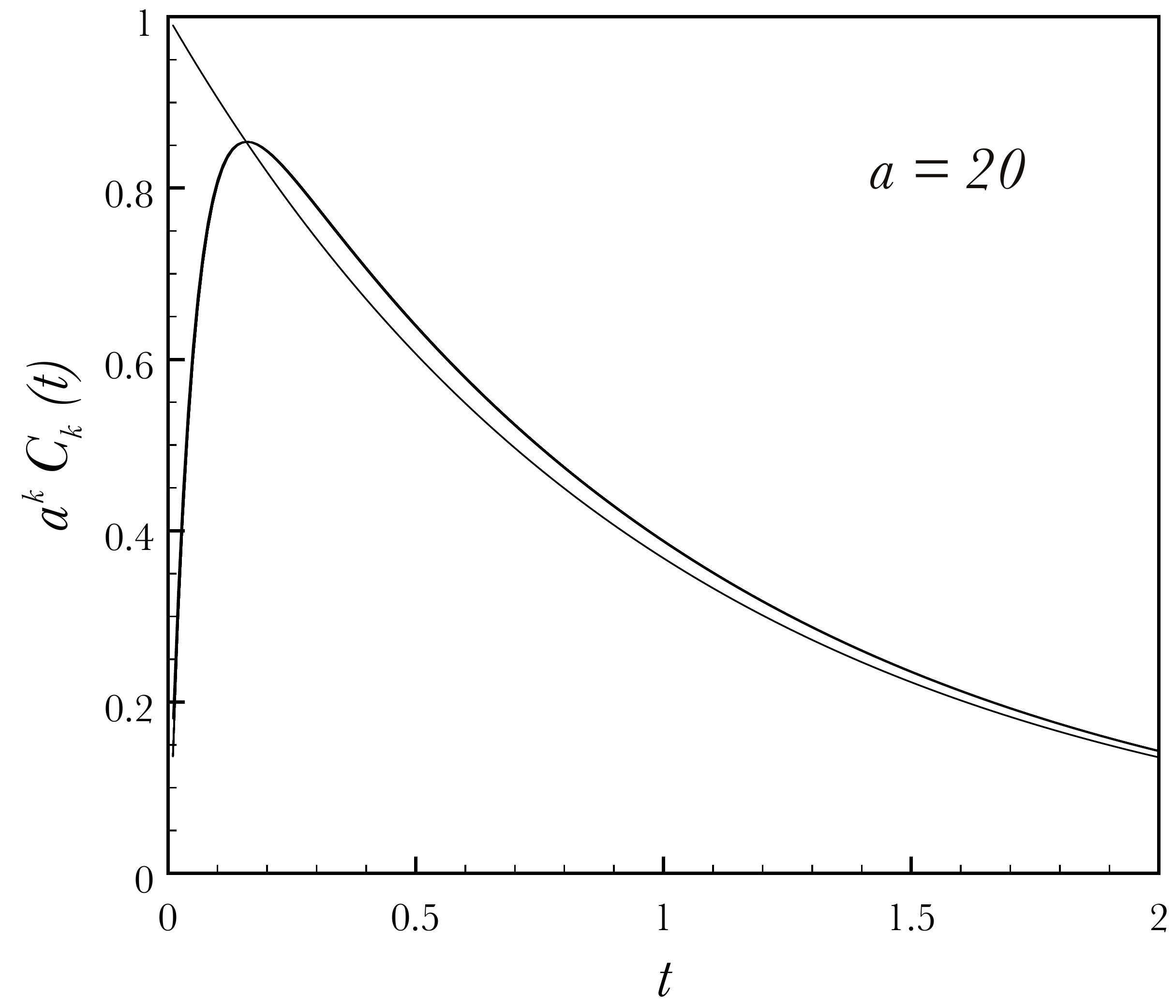} 
		\caption{The coefficients $a^kC_k(t)$ for increasing values of $a$, with $a=1.1, 2, 5,20$, and various values of $k$ ($k=0,\cdots, 10$ from left to right). These values of $k$ correspond to distinct curves only in the case $a=1.1$, where the behavior is close to that of model A ($a=1$), the energy in each mode reaching a maximum at time $t \sim k$. The flat behavior at small time that develops as $k$ grows clearly shows the delay in the excitation of modes with high $k$ values.  As $a$ grows, the curves corresponding to the various values of $k$ gradually merge together, the merging occurring from the largest $k$ values. Thus for instance for $a=2$, the curves corresponding to $k>4$ are indistinguishable from that representing $a^4 C_4(t)$, for $a=5$, only the curves representing $aC_1(t)$ and $a^2 C_2(t)$ are visible. As $a$ increases further, all curves merge onto that representing $aC_1(t)$, the latter  becoming itself  closer and closer to $C_0(t)=\rme^{-t}$, except at very small $t$. 
		}
		\label{limits}

\end{figure}

The other  interesting limit is that of large $a$, corresponding to a rapidly accelerating cascade. In this case, the numerical calculation  suggests that  the coefficients $ C_k(t)$ with $k\ge 1$ quickly reach the scaling behavior $C_k\sim 1/a^k$.    This is illustrated in Fig.~\ref{limits}, right panel. In fact when $a\gg 1$, and for not too small times,  $a^k C_k(t)\approx a C_1(t)$, and $a C_1(t)$ itself becomes closer and closer to $C_0(t)$ as $a$ grows, as can be inferred from the explicit expressions
\beq
C_1(t)=\frac{\rme^{- t}-\rme^{-a t}}{a-1},\qquad C_0(t)=\rme ^{-t}.
\eeq
 At very small $t$ ($at\ll 1$), however, the behavior is reminiscent of that found for $a=1$, 
\beq
C_k(t)=a^{k(k-1)/2}\,\frac{t^k}{k!} -  a^{k(k-1)/2}\frac{1-a^k}{1-a}\,\frac{t^{k+1}}{(k+1)!}+{\cal O}(t^{k+2}).
\eeq
The vanishing of the first $k$ coefficients of the Taylor expansion of the coefficient $C_k(\tau)$ may be seen as a non perturbative effect, corresponding to the fact that modes with high values of $k$ are not excited instantaneously. The delay stays however finite a $k$ grows, in contrast to what happens in  model A, where it grows linearly with $k$. These  behaviors of the coefficients $C_k$ in the two models can be see in Fig.~\ref{figb1} of the main text. 
A similar analysis can be done for the case with an external source of strength ${\cal A}$. Only the equation for the  coefficient $C_0$ depends explicitly on ${\cal A}$:
\beq\label{C0source}
C_0(t)={\cal A}\left(  1-\rme^{-t}   \right), \qquad
C_0(\omega)={\cal A}\left(  \frac{1}{\omega} -\frac{1}{1+\omega}\right).
\eeq
As compared to the case without source, Eq.~(\ref{CkLaplace}), $C_0(\omega)$  contains an additional pole at $\omega=0$ corresponding to the stationary source. Using the same method as  that used in the absence of source, one gets, for model B, 
\beq
\frac{1}{\cal A}\,C_{k\ge 1}(t)=\frac{1}{a^k}-\rme ^{-t} \prod_{m=1}^k \frac{a^{m-1}}{a^m-1}+\sum_{m=1}^k \frac{\rme^{-a^m t} }{a(a^m-1)} \prod_{n\ne m}^k \frac{a^{n - 1}}{a^n - a^m}.\nn
\eeq
 For model A ($a=1$), one recovers by this method the expression (\ref{CkAws}) of the main text. As was the case in the absence of the source, when $a\gg 1$,  $a^k C_k$ for $k\geq 1$ behaves essentially as $aC_1(t)$, and eventually as $C_0(t)$ for very large  $a$ and not too small $t$,  with 
  \beq
 C_1(t)={\cal A}  \left(\frac{1-\rme^{-a t}}{a}-\frac{\rme^{- t}-\rme^{-a t}}{a-1}\right),
  \eeq
  and $C_0(t)$ given by Eq.~(\ref{C0source}).
  At small $t$ one observes  the same  non perturbative behavior as in the case without the source, with $C_k(t)$ given by
  \beq
 \frac{1}{{\cal A}} C_k(t)=\frac{a^{k(k-1)/2}t^{k+1}}{(k+1)!}-\frac{t^{k+2}}{(k+1)!}\frac{1-a^{k+1}}{1-a}+{\cal O}(t^{k+3}).
 \eeq
 Again, this behavior is visible only in a very small time interval, $ta\ll 1$.
 
 %%%%%%%%%%%%%%%%%%%%%
\section{Exact solutions: constant kernel}\label{appconstkern}
%%%%%%%%%%%%%%%%%%%%%
In this section we provide the solutions of the two models A and B with a flat kernel
\beq
{\cal K}(z)=1.
\eeq

 %%%%%%%%%%%%%%%%%%%%%%%%%%%%%%%%%%
\subsection{Model A }\label{flat-A}
%%%%%%%%%%%%%%%%%%%%%%%%%%%%%%%%%%%
The evolution equation for model A ($\gamma=0$, $a=1$) reads
\beq\label{DGLAP-2}
\frac{\del}{\del t} D(x,t)= x\int_x^1 \frac{\rmd\xi}{\xi^2}D(\xi,t)-\frac{1}{2}D(x,t).
\eeq
In order to solve this equation, we perform first  the following change of variables, 
\beq
y=\ln\frac{1}{x}\,,\quad y'=\ln\frac{1}{\xi}\quad  \text{and}\quad \ D(x,t)= x \rme^{-\frac{1}{2}t} g(y,t).
\eeq
Equation~(\ref{DGLAP-2}) then yields 
\beq
\frac{\del}{\del t} g(y,t)=\int_0^y \rmd y' g(y',t), 
\eeq
which can be  solved by  Laplace transform. We obtain 
\beq
g(y,t)=\int_{-i\infty}^{i\infty} \frac{\rmd\nu}{2\pi i } \, \rme^{\nu y +t/\nu} =\sqrt{ \frac{t}{y}}I_1(2\sqrt{ty}),
\eeq
where $I_1(x)$ is the modified Bessel function of the first kind. 
In terms of the original variables, the energy distribution in the cascade reads
\beq
D(x,t)=x\sqrt{ \frac{t}{\ln\frac{1}{x}}}\, I_1\left( 2\sqrt{ t \ln\frac{1}{x}}\right)\,\rme^{-\frac{1}{2}t}. 
\eeq
When $x\ll \rme^{-t}$, the behavior of $D(x,t)$  is analogous to that of the DGLAP cascade (cf. Eq.~(\ref{DGLAPas0})), 
\beq\label{asympAflat}
D(x,t) \approx \left(\frac{1}{x}\right)^{\sqrt{\frac{2t}{\ln\frac{1}{x}}}-1}.
\eeq 
In particular, it vanishes at $x=0$, that is, an infrared cutoff is dynamically generated. This is enough to ensure that the energy of the cascade remains in the spectrum, as the following calculation shows
\beq
\int_0^1 dx D(x,t) &= & \rme^{-\frac{1}{2}t } \int_0^1 dx x \int \frac{\rmd\nu}{2\pi i } \, \rme^{\nu y +t/\nu}\,,\nn
&=&  \int \frac{d\nu}{2\pi i } \int_0^1 \rmd y   \, \rme^{(\nu-2) y +(1/\nu-1/2)t}\, \nn
&=&  \int \frac{\rmd\nu}{2\pi i } \frac{1}{\nu-2}\, \rme^{ (1 /\nu-1/2)t}\, \nn
&=&   \rme^{ (2 /\nu-1)t}\Big|_{\nu=2}\, \nn
&=& 1\,.
\eeq
In the problem with a source ${\cal A}\delta(1-x)$, the function  $g(y,t)$ becomes
\beq
g(y,t)={\cal A} \int \frac{\rmd\nu}{2\pi i } \,\frac{2\nu}{\nu-2}\left(\rme^{\frac{t}{2}}-\rme^{\frac{t}{\nu}}\right) \rme^{\nu y},
\eeq
which yields the steady state solution
\beq
D_\text{st}=\frac{4{\cal A}}{x}.
\eeq
Note that this  solution exists only for not too small $x$. At very small $x$, the dynamically generated cutoff suppresses the flow of energy, and $D(x,t)$ goes to zero.
%%%%%%%%%%%%%%%%%%%%%%%%%%%%%%%%%%
\subsection{Model B }\label{flat-B}
%%%%%%%%%%%%%%%%%%%%%%%%%%%%%%%%%%%
For model B ($\gamma=1/2$, $a=\sqrt{2}$), the transport equation  reads 
\beq\label{model-B-const}
\frac{\del}{\del t} D(x,t) = x\int_x^1 \frac{\rmd\xi}{\xi^{5/2}} D(\xi,t)-\frac{1}{2\sqrt{x}}D(x,t).
\eeq
This is to be solved with the initial condition $D_0(x)=\delta(1-x)$. 
To solve this equation, we start by shifting the distribution by the solution at the singular point $x=1$, 
\beq
D(x)= g(x)+\,\rme^{-\frac{1}{2}t }  \delta(1-x).
\eeq 
Eq.~(\ref{model-B-const}) then yields 
\beq\label{model-B-const-2}
\frac{\del}{\del t} g(x,t) =x\rme^{-\frac{1}{2}t }+  x\int_x^1 \frac{\rmd\xi}{\xi^{5/2}} g(\xi,t)-\frac{1}{2\sqrt{x}} g(x,t).
\eeq
If we use the ansatz $g(x)\sim x^{\alpha}$ in the r.h.s. of Eq.~(\ref{model-B-const-2}), we obtain
\beq\label{ansatz}
x\int_x^1 \frac{\rmd\xi}{\xi^{5/2}} \xi^\alpha-\frac{1}{2\sqrt{x}}x^\alpha=\frac{1}{2\alpha-3} \left(2x-\frac{1+2\alpha}{2}x^{\alpha-\frac{1}{2}}\right).\nn
\eeq
Imagine now solving Eq.~(\ref{model-B-const-2})  iteratively.  The leading order term is  $\sim x $, so that  in the first iteration $\alpha=1$, which yields a term proportional to $x$ and one proportional to $\sqrt{x}$. The next iteration will generate the powers  $x$, $\sqrt{x}$ and $x^0$,  the third iteration,  $x$, $\sqrt{x}$, $1/\sqrt{x}$ and $x^0$. In the fourth iteration, four powers are generated, $\alpha=-1/2,0,1/2,1$, but the $\alpha=-1/2$ contribution cancels with the second term in the r.h.s of Eq.~(\ref{ansatz}) that generates a $1/x$ term. It follows that the iterative procedure  does not generate other powers than $\alpha=-1/2,0,1/2,1$. 
As a consequence, the solution takes the general form
\beq\label{ansatz-general}
g(x,t)=f_{-1/2}(t)\, \frac{1}{ \sqrt{x} }+ f_0(t) +f_{1/2}(t) \,\sqrt{x}+f_{1}(t) \,x.
\eeq
Plugging Eq.~(\ref{ansatz-general}) into Eq.~(\ref{model-B-const-2}) and matching the coefficients of the four different powers of $x$ we obtain a closed system of equations, 
\beq\label{sys-eq}
\dot f_1 &=&\rme^{-\frac{1}{2}t } - 2f_1 - f_{1/2} - \frac{2}{3} f_0- \frac{1}{2} f_{-1/2},\nn
\dot f_{1/2}&=& \frac{3}{2} f_1 ,\nn
\dot f_{0}&=& \frac{1}{2} f_{1/2} ,\nn
\dot f_{-1/2}&=&\frac{1}{6} f_0.
\eeq
%Taking four  time  derivatives of the first equation yields an independent equation for $f_1$, 
%\beq
%\ddddot f_1 &=&-\frac{1}{8}e^{-\frac{1}{2}t } -2  \dddot f_1 -\frac{3}{2} \ddot f_{1} - \frac{1}{2} \dot f_1-\frac{1}{16} f_{1}.
%\eeq
At this point, it is convenient to perform a Laplace transform 
\beq
f_n(\nu)=\int_0^\infty \rmd t \rme^{-\nu t} f_n(t).
\eeq
Using the boundary conditions
\beq
f_n(\infty)=0\quad \text{and}\quad  f_n(0)=0,
\eeq
we then easily obtain
\beq\label{sys-eq-2}
\nu f_1 &=&\frac{1}{\nu+\frac{1}{2} } - 2f_1 - f_{1/2} - \frac{2}{3} f_0- \frac{1}{2} f_{-1/2},\nn
\nu f_{1/2}&=& \frac{3}{2} f_1 ,\nn
\nu f_{0}&=& \frac{1}{2} f_{1/2} ,\nn
\nu f_{-1/2}&=&\frac{1}{6} f_0.
\eeq
It is straightforward to solve this system of equations. We first get for $f_1$ (for $\nu\neq 0$)
\beq
\nu f_1(\nu)= \frac{1}{\nu+\frac{1}{2} } - 2f_1-\frac{3}{2 \nu} f_1- \frac{1}{2\nu^2} f_1- \frac{1}{16\nu^3} f_1,
\eeq
which yields
\beq
\left(\nu^4+2\nu^3+\frac{3}{2}\nu^2+\frac{1}{2}\nu+\frac{1}{16}\right)f_1(\nu)= \frac{\nu^3}{\nu+\frac{1}{2}}.
\eeq
It is easy to check that the polynomial in the l.h.s of the latter equation has a single zero, $\nu=-1/2$, so that
\beq
\nu^4+2\nu^3+\frac{3}{2}\nu^2+\frac{1}{2}\nu+\frac{1}{16}=\left(\nu+\frac{1}{2}\right)^4.
\eeq
We can then readily invert the Laplace transform, and get
\beq
f_1(t)&=&\int_{-i\infty}^\infty \frac{\rmd\nu}{2\pi i} \frac{\nu^3}{\left(\nu+\frac{1}{2}\right)^5} \rme^{\nu t}=\frac{1}{4! }\left(24 t-18t^2+3t^3-\frac{1}{8}t^4\right) \rme^{-\frac{1}{2}t}.\nn
f_{1/2}(t)&=&\frac{3}{2}\int_{-i\infty}^\infty \frac{\rmd\nu}{2\pi i} \frac{\nu^2}{\left(\nu+\frac{1}{2}\right)^5} \rme^{\nu t}=\frac{3}{2\times 4! }\left(12t^2-4t^3+\frac{1}{4}t^4\right) \rme^{-\frac{1}{2}t}.\nn
f_0(t)&=&\frac{3}{4} \int_{-i\infty}^\infty\frac{\rmd\nu}{2\pi i} \frac{\nu}{\left(\nu+\frac{1}{2}\right)^5} \rme^{\nu t}=\frac{3}{4\times 4! }\left(4t^3-\frac{1}{2}t^4\right) \rme^{-\frac{1}{2}t}.\nn
f_{-1/2}(t)&=&\frac{1}{8}\int_{-i\infty}^\infty \frac{\rmd\nu}{2\pi i} \frac{1}{\left(\nu+\frac{1}{2}\right)^5} \rme^{\nu t}= \frac{1}{8\times 4! }t^4 \rme^{-\frac{1}{2}t}.\nn
\eeq
For the case with a source, the equation reads
\beq\label{model-B-const-source}
\frac{\del}{\del t} D(x,t) ={\cal A} \delta(1-x)+ x\int_x^1 \frac{\rmd\xi}{\xi^{5/2}} D(\xi,t)-\frac{1}{2\sqrt{x}}D(x,t).
\eeq
Following the same steps as before, we look for a solution in the form
\beq
D(x,t)=g(x,t)+\left(1-\rme^{-\frac{1}{2}t}\right)\delta(1-x).
\eeq
The coefficient  $f_{-1/2}$ is given by
\beq
f_{-1/2}(t)&=&\frac{{\cal A}}{4}\int \frac{\rmd\nu}{2\pi i} \left(\frac{1}{\nu\left(\nu+\frac{1}{2}\right)^4}-\frac{1}{\left(\nu+\frac{1}{2}\right)^5} \right)\rme^{\nu t}.\nn
&=&\frac{{\cal A}}{4} \left[16 - \left(16+8t+2t^2+\frac{1}{3}t^3+\frac{1}{ 24 }t^4\right)  \rme^{-\frac{1}{2}t} \right].
\eeq
The remaining coefficients are then easily obtained
\beq
f_1(t)&=&{\cal A}\left(\frac{1}{2}t^2-\frac{1}{6}t^3+\frac{1}{96 }t^4\right) \rme^{-\frac{1}{2}t}.\nn
f_{1/2}(t)&=&{\cal A}\left(\frac{1}{4}t^3-\frac{1}{32 }t^4\right) \rme^{-\frac{1}{2}t}.\nn
f_0(t)&=&\frac{{\cal A}}{32 }t^4 \rme^{-\frac{1}{2}t}.\nn
\eeq
At late times, the constant contribution in the term $f_{-1/2}(t)$ dominates, and the solution becomes the stationary solution 
\beq
D_\text{sc}(x)=\frac{4{\cal A}}{\sqrt{x}}.
\eeq

%%%%%%%%%%%%%%%%%%%%%%%%%%%%%%%%%%%%

\end{document}